\documentclass[a4paper,11pt]{article}
\pdfoutput=1 

\usepackage{jheppub} 

\usepackage[T1]{fontenc} 

\usepackage{subfigure}
\usepackage{graphicx,amssymb,amsmath,amsthm,amsfonts,epsfig,epsf}
\usepackage[usenames]{color}
\definecolor{darkred}{rgb}{0.5,0,0}

\usepackage{amssymb,amsmath,amsthm}
\usepackage{mathrsfs}

\usepackage{array}
\usepackage{afterpage}
\usepackage{bm}
\usepackage{dcolumn}
\usepackage[latin1,utf8]{inputenc}
\usepackage{latexsym}
\usepackage{rotating}
\usepackage{longtable}

\usepackage{enumerate}
\usepackage{tensor,multirow}
\usepackage{url}
\usepackage{placeins}

\usepackage{float}
\usepackage{graphicx}

\newcounter{mnotecount}[section]

\renewcommand{\themnotecount}{\thesection.\arabic{mnotecount}}

\newcommand{\mnote}[1]
{\protect{\stepcounter{mnotecount}}$^{\mbox{\footnotesize
$
\bullet$\themnotecount}}$ \marginpar{
\raggedright\tiny\em
$\!\!\!\!\!\!\,\bullet$\themnotecount: #1} }

\title{\boldmath Strong Cosmic Censorship in Horndeski Theory}

\author[1]{Kyriakos Destounis,}
\author[2]{Rodrigo D. B. Fontana,}
\author[3,4]{Filipe C. Mena,}
\author[5]{Eleftherios Papantonopoulos,}

\affiliation[1]{CENTRA, Departamento de F\'{\i}sica, Instituto Superior T\'ecnico -- IST, Universidade de Lisboa -- UL,
Avenida Rovisco Pais 1, 1049 Lisboa, Portugal}
\affiliation[2]{Universidade Federal da Fronteira Sul, Campus Chapecó-SC
Rodovia SC 484 - Km 02, Fronteira Sul, CEP 89815-899, Brasil}
\affiliation[3]{Centro de An\'alise Matem\'atica, Geometria e Sistemas Din\^amicos, Instituto Superior T\'ecnico, Universidade de Lisboa, Avenida Rovisco Pais 1, 1049-001 Lisboa, Portugal}
\affiliation[4]{Centro de Matem\'atica, Universidade do Minho, 4710-057 Braga, Portugal}
\affiliation[5]{Physics Division, National Technical University of Athens, 15780 Zografou Campus, Athens, Greece}

\emailAdd{kyriakosdestounis@tecnico.ulisboa.pt}
\emailAdd{rodrigo.fontana@uffs.edu.br}
\emailAdd{filipecmena@tecnico.ulisboa.pt}
\emailAdd{lpapa@central.ntua.gr}

\abstract{The strong cosmic censorship hypothesis has recently regained a lot of attention in charged and rotating black holes immersed in de Sitter space. Although the picture seems to be clearly leaning towards the validity of the hypothesis in Kerr-de Sitter geometries, Reissner-Nordstr\"{o}m-de Sitter black holes appear to be serious counter-examples. Here, we perform another test to the hypothesis by using a scalar field perturbation non-minimally coupled to the Einstein tensor propagating on Reissner-Nordstr\"{o}m-de Sitter spacetimes. Such non-minimal derivative coupling is characteristic of Horndeski scalar-tensor theories. Although the introduction of higher-order derivative couplings in the energy-momentum tensor increases the regularity requirements for the existence of weak solutions beyond the Cauchy horizon, we are still able to find a small finite region in the black hole's parameter space where strong cosmic censorship is violated.}

\begin{document}
\maketitle
\flushbottom

\section{Introduction}

The theory of General Relativity (GR) is the most successful theory of gravitation. It predicts and properly describes black-hole (BH) spacetimes, gravitational-wave emission, cosmic expansion and many more phenomena. It is undoubtedly a cornerstone of modern theoretical physics and astronomy. One of the most fascinating attributes of GR is its ability to predict the future evolution of the spacetime and its matter constituents. Although this holds true for many solutions of the field equations, some BHs possess inner horizons beyond which the future evolution of the spacetime is highly non-unique.

The strong cosmic censorship (SCC) hypothesis \cite{Penrose69} conjectured by Penrose, states that appropriate initial data should be future inextendible beyond the Cauchy horizon (CH). Such a horizon designates the boundary of the maximal (globally hyperbolic) development of initial data beyond which the Einstein field equations lose their predictive power. Therefore, for GR to maintain its deterministic nature, SCC should be respected in BH spacetimes and, in the more modern and precise formulation of Christodoulou \cite{Christodoulou:2008nj}, this requires that the spacetime metric should be inextendible beyond the CH, even as a weak solution of the field equations.

Recent studies indicate that Reissner-Nordstr\"om-de Sitter (RNdS) geometries might violate SCC. The work of \cite{Cardoso:2017soq} has provided strong numerical evidence which indicate that perturbed RNdS BHs are serious counter-examples to the SCC conjecture. Such prediction generalizes to RNdS BHs with a back-reacting scalar field \cite{Luna:2018jfk}. Therefore, such spherically-symmetric electrically-charged solutions with a positive cosmological constant do not respect SCC.

Most linearized studies of RNdS are based on two fundamental phenomena: (i) the exponential blueshift effect at the CH, which might lead to the blow-up of the energy density \cite{Dafermos:2003wr,Dafermos:2012np} and to a mass-inflation singularity \cite{BradyPoisson,Ori:1991zz} and (ii) the exponential decay of perturbations on the exterior of asymptotically de Sitter (dS) spacetimes \cite{Hintz:2016gwb,Hintz:2016jak}. These two phenomena can compete and, in some cases, counter-balance each other leading to a CH with enough regularity to allow the spacetime metric and the matter field to be extended past it into a region where the field equations cannot predict, unambiguously, the evolution of initial data. As a matter of fact, the spacetime and matter can be extended as weak solutions to the Einstein equations, thus violating Christodoulou's formulation of SCC.\footnote{In the case of asymptotically flat spacetimes, perturbations decay polynomially in the exterior and therefore cannot compete with the exponential blueshift at the CH and, thus, SCC is respected.}

Another key ingredient is the blueshift and redshift factors that govern the growth and decay of perturbations in the interior and exterior of RNdS BHs, respectively. The blueshift amplification is governed by the surface gravity \cite{Hartle} of the CH, $\kappa_-$, while the redshift decay is governed by the spectral gap, $\alpha$, which corresponds to the imaginary part of the dominant, non-zero, quasinormal mode (QNM) \cite{Hintz:2016gwb,Hintz:2016jak} (for a review on QNMs see \cite{Kokkotas:1999bd,Berti:2009kk,Konoplya:2011qq}). The combination of these ingredients in Einstein-Maxwell-scalar field theories leads to the definition of a control parameter, $\beta\equiv\alpha/\kappa_-$, which decides the fate of SCC \cite{Hintz:2015jkj}, in the sense that if $\beta>1/2$, then SCC is violated.

In \cite{Cardoso:2017soq}, it was shown that near-extremally charged Reissner-Nordstr\"om-de Sitter BHs, lead to possible violations of SCC under neutral massless scalar perturbations. This effect can become even more severe in the case of the coupled electromagnetic and gravitational perturbations \cite{Dias:2018etb}. Although it has been argued that a charged scalar field is enough to preserve the validity of SCC  \cite{Hod:2018dpx,Hod:2018lmi}, there is still a finite volume of the parameter space of near-extremal RNdS BHs where SCC may be violated \cite{Cardoso2,Zhang1,Dias:2018ufh}. This region is small, but still existent, due to a superradiant instability occurring in RNdS when spherically-symmetric charged scalar fields are scattered off the BH \cite{Zhu:2014sya,Konoplya:2014lha,Destounis:2019hca}. Charged Dirac field perturbations in RNdS were also inadequate to prevent the violation of SCC \cite{Ge:2018vjq,Destounis:2018qnb}. Although spherically-symmetric spacetimes look problematic in the context of SCC, the same does not occur for rotating geometries. In \cite{Dias:2018ynt}, scalar and gravitational perturbations of Kerr-de Sitter (KdS) seem to respect the linearized analogue of SCC, although the opposite seems to happen for Dirac perturbations \cite{Rahman:2019uwf}. An interesting study of the dimensional influence on the validity of SCC in higher-dimensional RNdS and KdS was carried out in \cite{Liu:2019lon,Rahman:2018oso}. In particular, it was shown that higher-dimensional RNdS BHs may still violate SCC under scalar perturbations, however higher-dimensional KdS BHs do not.

On the mathematical analysis side, an interesting suggestion to restore SCC in RNdS BHs, was proposed in \cite{Dafermos:2018tha}, where it was shown that the pathologies identified in \cite{Cardoso:2017soq} become non-generic if one considerably enlarges the allowed set of initial data by weakening their regularity. The considered data are also compatible with Christodoulou's formulation of SCC.

Further linearized studies of SCC were carried out in theories in which the scalar field was not minimally coupled to the curvature. In \cite{Gwak1}, SCC was studied under a non-minimally coupled massive scalar field on lukewarm RNdS and MTZ BHs. There, it was argued that the validity of the SCC conjecture depends on the scalar field properties. More recently, in \cite{Guo:2019tjy}, SCC was investigated in RNdS by examining the evolution of a scalar field non-minimally coupled to the Ricci curvature. It was found that the stability of the BH background and the fate of SCC depend on the coupling of the scalar field to curvature. Furthermore, Born-Infeld-dS spacetimes seem to possess similar problems as the ones depicted in near-extremal RNdS \cite{Gan:2019jac}.

Considering this debate, and the importance of the validity of SCC, it would be interesting to further analyze whether the violations that occur in GR are still present in modified theories of gravity and if they tend to vanish, in some sense, after the inclusion of corrections to GR.

Gravitational theories that modify the standard Einstein theory of gravity have a long history. One  recent class of these theories is the Horndeski scalar-tensor theory \cite{Horndeski1974} which involves second order field equations in four
dimensions \cite{Nicolis:2008in,Deffayet:2009wt,Deffayet:2009mn}. One important term appearing in the Horndeski Lagrangian is
the kinetic coupling  of a scalar field to the Einstein tensor. This term provides important modifications to the standard gravity theory with a minimally coupled scalar field in both small and large distances (for a review of this effect see \cite{Papantonopoulos:2019eff}).

The main effect of  the kinetic coupling of a scalar field to the Einstein tensor is that it influences strongly the kinetic properties of the scalar field, acting as a friction term. This was observed in cosmology \cite{Sushkov:2009hk,Germani:2010gm} and in local BH solutions, where it was found that when the coupling becomes stronger, it takes more time for a BH to form \cite{Koutsoumbas:2015ekk}. The stability of BHs in scalar-tensor theories in the presence of this coupling was discussed in \cite{Kolyvaris:2017efz,Kolyvaris:2018zxl}.

The QNMs of a BH with a scalar field coupled to the Einstein tensor were calculated in \cite{Minamitsuji:2014hha}. Furthermore, the QNMs for a class of static and spherically symmetric BH
containing the derivative coupling were studied in \cite{Yu:2018zqd}. In turn, calculations of QNMs for a massive scalar field,
with the derivative coupling on a RN background, were performed in \cite{Konoplya:2018qov}, while vectorial and spinorial perturbations in Galileon BHs were performed in \cite{Abdalla:2018ggo}. The effects of the coupling of a scalar field to the Einstein tensor on the stability of RNdS \cite{Fontana:2018fof} and RNAdS BHs \cite{Abdalla:2019irr} were investigated very recently. These studies indicate that the decay of perturbations is strongly influenced by an increasing non-minimal coupling and, in some cases, lead to the destabilization of the BH exterior.

The motivation of our work is threefold: (i) Do perturbations still decay exponentially in Horndeski theory for asymptotically dS BHs and, if so, are they still dominated by the dominant QNMs at late times? (ii) Is $\beta$, still, supposed to be bounded by $1/2$ after the introduction of higher-order derivative coupling terms to the energy-momentum tensor and, if not, what is the bound beyond which SCC is violated? (iii) Is SCC still violated by scalar fields non-minimally coupled to the Einstein tensor in RNdS spacetimes?

In the following sections, we aim to answer all these questions in the context of a particular Horndeski theory. We will show that scalar fields non-minimally coupled to the Einstein tensor, decay exponentially in the exterior of RNdS BHs and the late-time behavior is still dominated by the longest-lived QNM. We will prove that, in the particular Horndeski theory, $\beta>1/2$ is not adequate to decide the fate of SCC. In fact, we will show that the existence of weak solutions beyond the CH of RNdS spacetime and the violation of SCC occurs when $\beta>3/2$. Finally, we will demonstrate that, even though the regularity requirement for violation is considerably higher in the specific theory, ``small'' RNdS BHs still violate SCC for sufficiently large non-minimal Horndeski couplings $\eta$, which satisfy $|\eta|<1$. By considering such coupling range, we will not deviate significantly from GR and we will avoid potential discontinuities and instabilities.

\section{Weak solutions to the Einstein equations in Horndeski theory}
To study the extendibility of solutions to the field equations beyond the CH of a BH spacetime, we will consider a neutral massless scalar field $\phi$ non-minimally coupled to the Einstein tensor, with coupling strength $\eta\in\mathbb{R}$, and the following action:
\begin{equation}
\label{action}
S=\int_\mathcal{M} d^4x \sqrt{-g}\left(\frac{\mathcal{R}-2\Lambda}{16\pi}-\frac{1}{4}F_{\mu\nu}F^{\mu\nu}-\frac{1}{2}\left(g^{\mu\nu}+\eta G^{\mu\nu}\right)\partial_\mu \phi\partial_\nu \phi\right),
\end{equation}
where we use units such that $c=G=1$,
$\mathcal{R}$ is the Ricci scalar, $F^{\mu\nu}\equiv \partial^\mu A^\nu-\partial^\nu A^\mu$ and $G^{\mu\nu}\equiv \mathcal{R}^{\mu\nu}-\frac{1}{2}g^{\mu\nu}\mathcal{R}$ are the electromagnetic and Einstein tensors, respectively, $A^\mu$ is the electromagnetic potential, $\mathcal{R}^{\mu\nu}$ the Ricci tensor and $\Lambda>0$ the cosmological constant. The spacetime solution that interests us is a spherically symmetric, electrically charged BH immersed in a Universe with a positive cosmological constant, namely the RNdS solution of the Einstein-Maxwell theory. Such a BH possess a CH in its interior  and is described by the line element
\begin{align}
\label{metric}
ds^2 = -f(r) dt^2+\frac{1}{f(r)}dr^2+r^2 d\Omega^2_2, \hspace{1.5cm} f(r)= 1-\frac{2M}{r}+\frac{Q^2}{r^2} - \frac{\Lambda r^2}{3},
\end{align}
with $M$, $Q$ the mass and charge of the BH and $d\Omega^2_2$ the metric on the 2-sphere. In the next section, we will discuss this spacetime in more detail. In this section, though, we keep the notation as general as possible although, in some steps, we use properties specific to the spherically symmetric static spacetimes \eqref{metric}, such as a metric dependent on the radial coordinate only.

We recall that varying \eqref{action} with respect to $\phi$ leads to the equation of motion
\begin{equation}
\label{KG}
\frac{1}{\sqrt{-g}}\partial_\mu\left(\sqrt{-g}\left(g^{\mu\nu}+\eta G^{\mu\nu}\right)\partial_\nu\phi\right)=0,
\end{equation}
while varying with respect to $F$ leads to Maxwell's equations
\begin{equation}
dF=d\star F=0,
\end{equation}
where $\star$ is the Hodge star operator. Finally, varying \eqref{action} with respect to $g_{\mu\nu}$ leads to the field equations
\begin{equation}
\label{field equation}
G_{\mu\nu}+\Lambda g_{\mu\nu}=8\pi T_{\mu\nu},
\end{equation}
where
\begin{equation}
\label{energy momentum}
T_{\mu\nu}=T_{\mu\nu}^{(s)}+T_{\mu\nu}^{(em)}+\eta\Theta_{\mu\nu},
\end{equation}
is the energy-momentum tensor associated with the scalar field $\phi$ coupled to the Einstein tensor and the electromagnetic tensor $F^{\mu\nu}$. $T_{\mu\nu}$ is divided in three parts: $T_{\mu\nu}^{(s)}$ for the scalar field, $T_{\mu\nu}^{(em)}$ for the electromagnetic field and $\Theta_{\mu\nu}$ for the higher-order derivative terms, as:
\begin{align}
\label{scalar}
T_{\mu\nu}^{(s)}&=\partial_\mu\phi\partial_\nu\phi-\frac{1}{2}g_{\mu\nu}\partial_\alpha\phi\partial^\alpha\phi,\\
\label{electromagnetic}
T_{\mu\nu}^{(em)}&=F_{~\mu}^\alpha F_{\mu\alpha}-\frac{1}{4}g_{\mu\nu}F_{\alpha\beta}F^{\alpha\beta},\\
\nonumber
\Theta_{\mu\nu}&=-\frac{1}{2}\partial_\mu\phi\partial_\nu\phi R+2\partial_\alpha\phi\partial_{(\mu}\phi R^\alpha_{~\nu)}-\frac{1}{2}G_{\mu\nu}\left(\partial\phi\right)^2+\nabla^\alpha\phi\nabla^\beta\phi R_{\mu \alpha\nu \beta}+\nabla_\mu\nabla^\alpha\phi \nabla_\nu\nabla_\alpha\phi\\
\label{higher order}&\,\,\,\,\,\,\,-\nabla_\mu\nabla_\nu\phi\Box \phi
+\frac{1}{2}g_{\mu\nu}\left[-\nabla^\alpha\nabla^\beta\phi\nabla_\alpha\nabla_\beta\phi+(\Box\phi)^2-2\partial_\alpha\phi\partial_\beta\phi R^{\alpha\beta}\right].
\end{align}
If we assume that $\phi$ and $g_{\mu\nu}$ are not necessarily $C^2$, i.e. twice continuously differentiable, we can still make sense of (\ref{field equation}) by multiplying  with a smooth, compactly supported, test function $\psi$ and integrating on both sides of the equation, in a small neighborhood $\mathcal{V}\subset \mathcal{M}$. If the outcome of the integral is bounded, then we can get a weak solution to \eqref{field equation}. Therefore, to have a weak solution at the CH, we require finiteness of
\begin{equation}
\label{terms}
\int_\mathcal{V}d^4x\sqrt{-g}(G_{\mu\nu}+\Lambda g_{\mu\nu}-8\pi T_{\mu\nu})\psi=0.
\end{equation}
The first two terms of \eqref{terms} are the usual ones which lead to the requirement of square integrability of the Christoffel symbols as follows:
\begin{align}
\nonumber
\int_\mathcal{V}d^4x\sqrt{-g}(G_{\mu\nu}&+\Lambda g_{\mu\nu})\psi\sim \int_\mathcal{V}d^4x\sqrt{-g}(\partial\Gamma+\Gamma^2+\Lambda g_{\mu\nu})\psi\\\label{Gmn}&\sim-\int_\mathcal{V} d^4x\sqrt{-g}(\partial\psi)\Gamma+\int_\mathcal{V}d^4x\sqrt{-g} \Gamma^2\psi+\Lambda\int_\mathcal{V}d^4x\sqrt{-g} g_{\mu\nu}\psi,
\end{align}
where we schematically expanded $G_{\mu\nu}\sim\Gamma^2+\partial\Gamma$, with $\Gamma$ denoting the Christoffel symbols, and we omit most of the indices. Therefore, for \eqref{Gmn} to be bounded, we require $\Gamma\in L^2_\text{loc}$, where $L^2_\text{loc}$ denotes the space of locally square integrable functions in $\mathcal V$. The third term of \eqref{terms} is the one that will define the higher regularity requirement, since higher-order derivative couplings are present in $T_{\mu\nu}$.

Let us consider first the standard energy-momentum tensor of a minimally coupled scalar field \eqref{scalar} (i.e. for $\eta=0$) schematically as:
\begin{equation}
\label{etazero}
\int_\mathcal{V} d^4x\sqrt{-g}T^{(s)}_{\mu\nu}\psi\sim\int_\mathcal{V}d^4x\sqrt{-g} (\partial\phi)^2\,\psi.
\end{equation}
This terms leads to the requirement of integrability of $(\partial\phi)^2$ or equivalently $\phi\in H^1_\text{loc}$, where $H^p_\text{loc}$ denotes the Sobolev space of functions in $L^2_\text{loc}$ such that their derivatives up to order $p$, in a weak sense, are also in $L^2_\text{loc}$.

The electrostatic potential $A_\mu=-\delta^0_{~\mu} Q/r$ associated with $F_{\mu\nu}$, sourced by the BH's charge $Q$, is regular at the CH and, therefore, \eqref{electromagnetic} does not further contribute to the regularity requirements.

By considering the terms of $\Theta_{\mu\nu}$ with higher-order derivative couplings, we realize that the first, second, fourth and ninth term of \eqref{higher order} yield integrability requirements through the following term
\begin{equation}
\label{coupled1}
\int_\mathcal{V} d^4x\sqrt{-g}(\partial\phi)^2 \mathcal{R}\, \psi\lesssim\sup(\psi)\int_\mathcal{V}d^4x\sqrt{-g}\left[(\partial\phi)^4+\mathcal{R}^2\right].
\end{equation}
The first term of \eqref{coupled1} requires integrability of the gradient of $\phi$ to the fourth power, while the second one has the form
\begin{align}
\nonumber
\int_\mathcal{V} d^4x\sqrt{-g}\mathcal{R}^2\sim\int_\mathcal{V}d^4x\sqrt{-g} (\partial\Gamma+\Gamma^2)^2&\sim\int_\mathcal{V}d^4x\sqrt{-g}\left[(\partial \Gamma)^2+2\Gamma^2\partial\Gamma+\Gamma^4\right]\\\label{R2}&\lesssim 2\int_\mathcal{V}d^4x\sqrt{-g} \left[(\partial^{2} g_{\mu\nu})^2+(\partial g_{\mu\nu})^4\right],
\end{align}
where, again, we schematically expanded $\mathcal{R}\sim\Gamma^2+\partial\Gamma$. For \eqref{R2} to be bounded, we require integrability of the gradient of $g_{\mu\nu}$ to the fourth power, or equivalently $\Gamma\in L^4_{\text{loc}}$, plus integrability of $(\partial^2 g_{\mu\nu})^2$ or equivalently $g_{\mu\nu}\in H^2_\text{loc}$. Following similar procedures, we realize that the third term of \eqref{higher order} requires
integrability of the gradient of $\phi$ to the fourth power and the remaining terms of \eqref{higher order} require finiteness of\footnote{For simplicity, we only demonstrate one of the terms involving second derivatives squared, but the rest lead to equivalent regularity requirements.}
\begin{equation}
\label{key}
\int_\mathcal{V} d^4x\sqrt{-g}(\Box\phi)^2\psi\sim\int_\mathcal{V}d^4x\sqrt{-g}(\partial^2\phi)^2\psi.
\end{equation}
This occurs if $(\partial^2\phi)^2$ is integrable at the CH, or equivalently $\phi\in H^2_\text{loc}$ (recall that for the case of $\eta=0$ the extra terms of $\Theta_{\mu\nu}$ would vanish and one only requires $\phi\in H^1_\text{loc}$ in accordance with \eqref{etazero} and \cite{Cardoso:2017soq}). By realizing that the scalar field and spacetime metric share similar regularity requirements \cite{Hintz:2016gwb,Hintz:2016jak,Costa:2017tjc,Dafermos:2017dbw}, it seems adequate to examine the behavior of $\phi$ at the CH.

We note that, in our framework, the mass is at the level of a spacetime integral of $T_{\mu\nu}$ regarding regularity \cite{Wald:1984rg,Faraoni:2015sja,Szabados2009}, so we do not not expect further restrictions on $\beta$ coming from requiring finiteness of the mass.

It is worth mentioning that if a Horndeski theory does not admit a well-posed initial value problem (IVP), then this would constitute a more severe problem than SCC itself. In \cite{Papallo:2017qvl,Papallo:2017ddx}, it is stated that the IVP expressed in the generalized harmonic gauge, for some Horndeski theories, is not strongly hyperbolic, although there could be a different choice of gauge in which the theories are strongly hyperbolic. Different methods, such as the ADM decomposition utilized in \cite{LeFloch:2014zva}, where the well-posedness of $f(\mathcal{R})$ theories is rigorously proven, could be considered to investigate such problem.
%

\section{Quasinormal modes of scalar fields coupled to the Einstein tensor in Reissner-Nordstr\"om-de Sitter spacetimes}

The causal structure of RNdS spacetime possesses three horizons, the CH at $r=r_-$, the event horizon at $r=r_+$ and the cosmological horizon at $r=r_c$. The horizons radii satisfy $r_-<r_+<r_c$ and the surface gravity of each horizon is given by:
\begin{equation}
\kappa_i=\Big|\frac{f^\prime(r)}{2}\Big|_{r=r_i},\,\,\,\,\,\,\,\,\,i\in\{-,+,c\}.
\end{equation}
The scalar field will be treated as a small perturbation which does not back-react to the fixed RNdS geometry for small $\eta$ (see e.g. \cite{Konoplya:2018qov,Fontana:2018fof}). To recast Eq. \eqref{KG} into a Schr\"odinger-like form, we take advantage of the symmetries of the spacetime and expand $\phi$ as
\begin{align}
\label{eqm3}
\phi\sim\sum_l\sum_{m}\frac{R(r,t)}{b(r)} \ Y_{l m}(\theta ,\phi ),
\end{align}
with $l,\,m$ the usual spherical harmonic indices, $b(r) = r\sqrt{k}$ and $k=1-\eta \left(\Lambda + {Q^2}/{r^4} \right)$. Therefore, the scalar field obeys the usual linear non-homogeneous wave equation
\begin{align}
\label{eqmstar}
\frac{\partial^2 R(r,t)}{\partial r_{*}^2} -\frac{\partial^2 R(r,t)}{\partial t^2}- V(r)R(r,t)=0,
\end{align}
with $dr_*=dr/f(r)$ defining the tortoise radial coordinate $r_*$ and $V(r)$ the effective potential
\begin{eqnarray}
\label{potential}
V(r)=f(r)\left(f(r)\frac{2rkk^{\prime\prime}+4kk^\prime-r(k^\prime)^2}{4rk^2}+f^\prime(r)\frac{2k+rk^\prime}{2rk}+\left(1+\frac{2Q^2\eta}{r^4 k}\right)\frac{l(l+1)}{r^2}\right).
\end{eqnarray}
The field expansion introduced in \eqref{eqm3}, transforms the equation of motion of $\phi$ to the usual Schr\"{o}dinger-like form \eqref{eqmstar} with a drawback: it introduces a discontinuity regime on $r$, whenever $b(r)=0$. To avoid such discontinuity we treat cases in which $\eta$ is small enough, considering the Horndeski action as a perturbative effect to GR. Such a choice is essential in order to consider the RNdS spacetime as a solution of GR. Therefore, by restricting to couplings satisfying $|\eta|<1$ we evade discontinuities in \eqref{potential} and subsequently  scalar field instabilities, as analyzed in \cite{Fontana:2018fof}.

If we assume a harmonic time dependence of the form $R(r,t)=\Phi(r) e^{-i\omega t}$, then \eqref{eqmstar} acquires the standard form
\begin{equation}
\label{master_eq}
\frac{d^2 \Phi}{d r_{*}^2}+(\omega^2-V(r))\Phi=0.
\end{equation}

To calculate the QNMs $\omega$ of \eqref{master_eq} we impose the following, physically motivated, boundary conditions:
\begin{equation}
\label{bcs}
\Phi \sim
\left\{
\begin{array}{lcl}
e^{-i \omega r_* },\,\,\,\quad r \rightarrow r_+, \\
&
&
\\
 e^{i\omega r_*},\,\,\,\,\,\,\quad r \rightarrow r_c.
\end{array}
\right.
\end{equation}

To extract the regularity requirement of $\phi$ at the CH of RNdS, we look at the asymptotic behavior of scalar waves there. As $r\rightarrow r_-$, then $V(r)\rightarrow 0$, so the first independent mode solution of \eqref{KG} can be expressed as
\begin{equation}
{\phi}_1\sim e^{-i\omega(t+r_*)},
\end{equation}
and the second as
\begin{equation}
{\phi}_2\sim e^{-i\omega(t-r_*)}.
\end{equation}
It is more convenient to use outgoing Eddington-Finkelstein coordinates, with $u=t-r_*$, which are regular at the CH. In that case
\begin{align}
\label{first}
{\phi}_1&\sim e^{-i\omega(t+r_*)}=e^{-i\omega u}e^{-2i\omega r_*},\\
\label{second}
\phi_2&\sim e^{-i\omega(t-r_*)}=e^{-i\omega u}.
\end{align}
Near the CH, the tortoise coordinate becomes
\begin{equation}
r_*=\int f^{-1}dr\sim \frac{\log|r-r_-|}{f^\prime(r_-)},
\end{equation}
where $f(r\rightarrow r_-)\sim|r-r_-|$ modulo irrelevant terms. Obviously, (\ref{second}) is regular, since for $r= r_- \,\,(u=\text{const.})$, ${\phi}_2$ is approximately constant. The solution which might introduce non-smoothness and, thus, defines the regularity requirement at the CH will be
\begin{equation}
{\phi}_1\sim e^{-i\omega u}e^{-2i\omega \log|r-r_-|/f^\prime(r_-)}=e^{-i\omega u} |r-r_-|^{-2i\omega/f^\prime(r_-)}=e^{-i\omega u}|r-r_-|^{i\omega/\kappa_-}.
\end{equation}
If we assume modes of the form $\omega=\omega_R+i\omega_I$, then
\begin{equation}
\label{mode behavior}
{\phi}_1\sim |r-r_-|^{i\omega_R/\kappa_-} |r-r_-|^{-\omega_I/\kappa_-}\sim |r-r_-|^{i\omega_R/\kappa_-} |r-r_-|^\beta,
\end{equation}
where we ignored the $e^{-i\omega u}$ factor, since it is smooth at the CH. The first factor of \eqref{mode behavior} is purely oscillatory, thus only the second factor plays a role in the asymptotic behavior of the scalar field. We recall that we have set $\beta\equiv\alpha/\kappa_-$, where $\alpha\equiv-\text{Im}(\omega)$ is the spectral gap or the imaginary part of the dominant, non-zero, QNM.

The highest requirement for the existence of extensions as weak solutions in the particular theory in study is $\phi\in H^2_\text{loc}$. Since the only relevant coordinate is the radial one, this translates to
\begin{equation}
\label{requirement2}
\int_\mathcal{V} (\partial_r^2 {\phi})^2dr\sim\int_\mathcal{V} |r-r_-|^{2(\beta-2)}dr\sim\frac{|r-r_-|^{2\beta-3}}{2\beta-3}.
\end{equation}
So, for \eqref{requirement2} to be finite at $r= r_-$, we require
\begin{equation}
\label{final}
\beta>\frac{3}{2}.
\end{equation}
Hence, in our theory, the requirement \eqref{final} has to be satisfied for the field equations \eqref{terms} to make sense in a weak manner at the CH. Recall that if $\eta=0$ then $\phi\in H^1_\text{loc}$ is enough to guarantee the violation of SCC, which leads to $\beta>1/2$, in accordance with \eqref{etazero} and \cite{Cardoso:2017soq}. If \eqref{final} holds, then the SCC hypothesis is violated. If, on the other hand, $\beta<3/2$ then at least the term \eqref{requirement2} will blow up and, thus, \eqref{terms} will be infinite. In that case, the field equations will not have weak solutions at $\mathcal{V}$ and SCC will be respected.

This novel threshold value for $\beta$ is clearly higher than the one found in previous studies. This is due to the existence of higher-order derivative terms that the non-minimal coupling to the Einstein tensor introduces to the energy momentum tensor. As a result, in the particular Horndeski theory, we require $\phi\in H^2_\text{loc}$ in order to have weak solutions of the field equations, whereas in previous studies the requirement was only $\phi\in H^1_\text{loc}$.

To calculate $\beta$ we need to extract the dominant QNMs of the non-minimally coupled scalar field propagating on a fixed RNdS background. We use two different numerical methods to do so. The first one relies on the {\it Mathematica} package {\it QNMSpectral} developed in \cite{Jansen:2017oag} (which, in turn, is based on methods proposed in \cite{Dias:2015nua}). The second one is based on the numerical integration of \eqref{eqm3} in the time domain, developed in \cite{Gundlach:1993tp}, where the wave equation is integrated in double-null coordinates, using two Gaussian wave-packets as Cauchy data. By applying the Prony method on the numerical evolution of the perturbation, we can then extract the QNMs. The Prony method was first introduced to the extraction of QNMs in \cite{Berti:2007dg}. In Appendix \ref{appA}, we demonstrate that both methods lead to QNMs which agree with very good precision in the parameter space region of interest (see Table \ref{table}). Furthermore, the trustworthy method of Wentzel-Kramers-Brillouin (WKB) approximation \cite{Schutz:1985zz} is used in various cases to justify the validity of our numerics at the eikonal limit.

\section{Dominant quasinormal-mode families of Reissner-Nordstr\"om-de Sitter black holes in Horndeski theory}

According to \cite{Cardoso:2017soq}, the region of interest in RNdS, where the violation of SCC may occur, lies close to charge extremality. With the aforementioned numerics, in the region of interest, we find, again, three distinct families of modes:
\begin{figure}[t]
\centering
\includegraphics[scale=0.2]{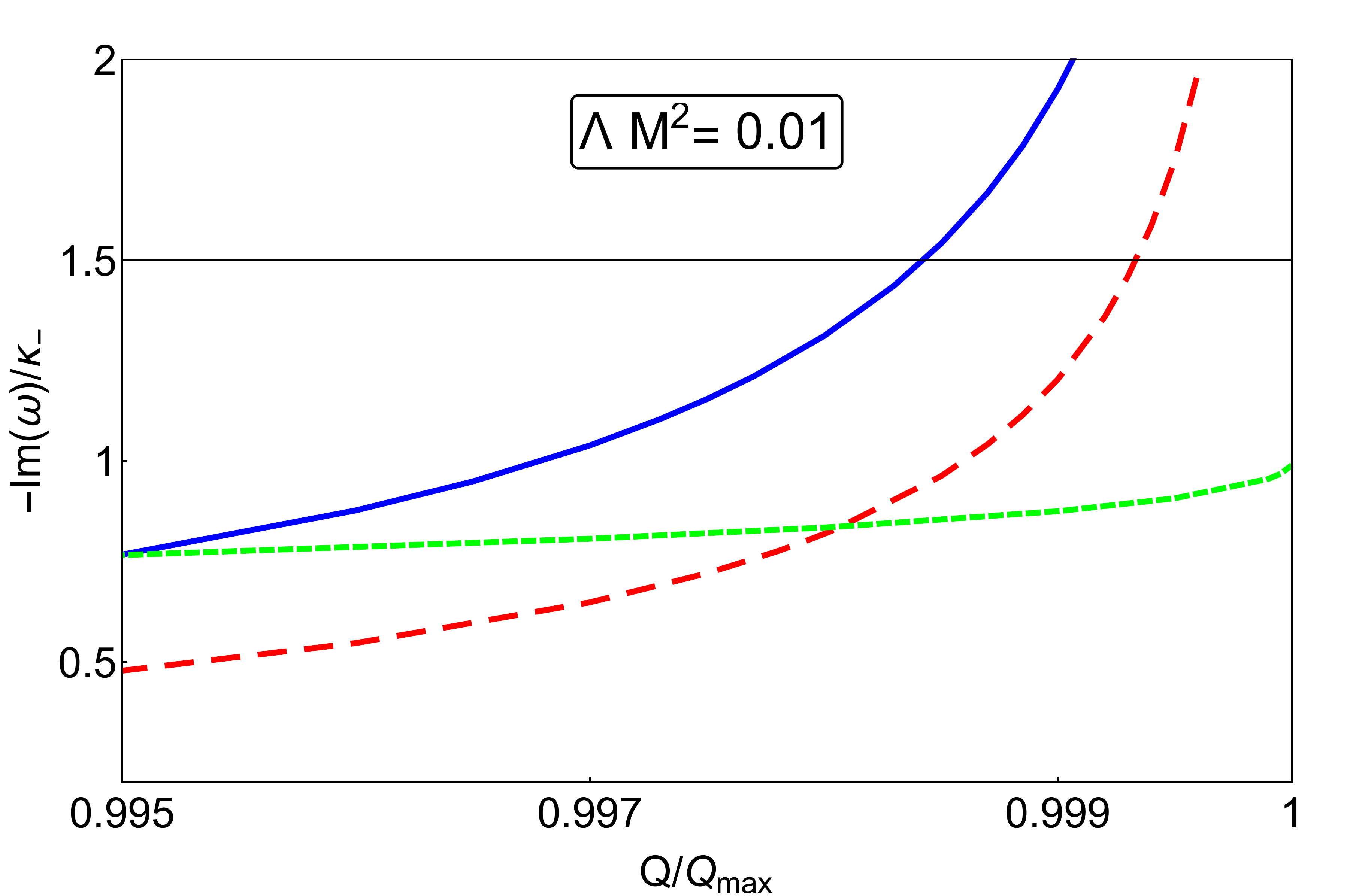}
\includegraphics[scale=0.2]{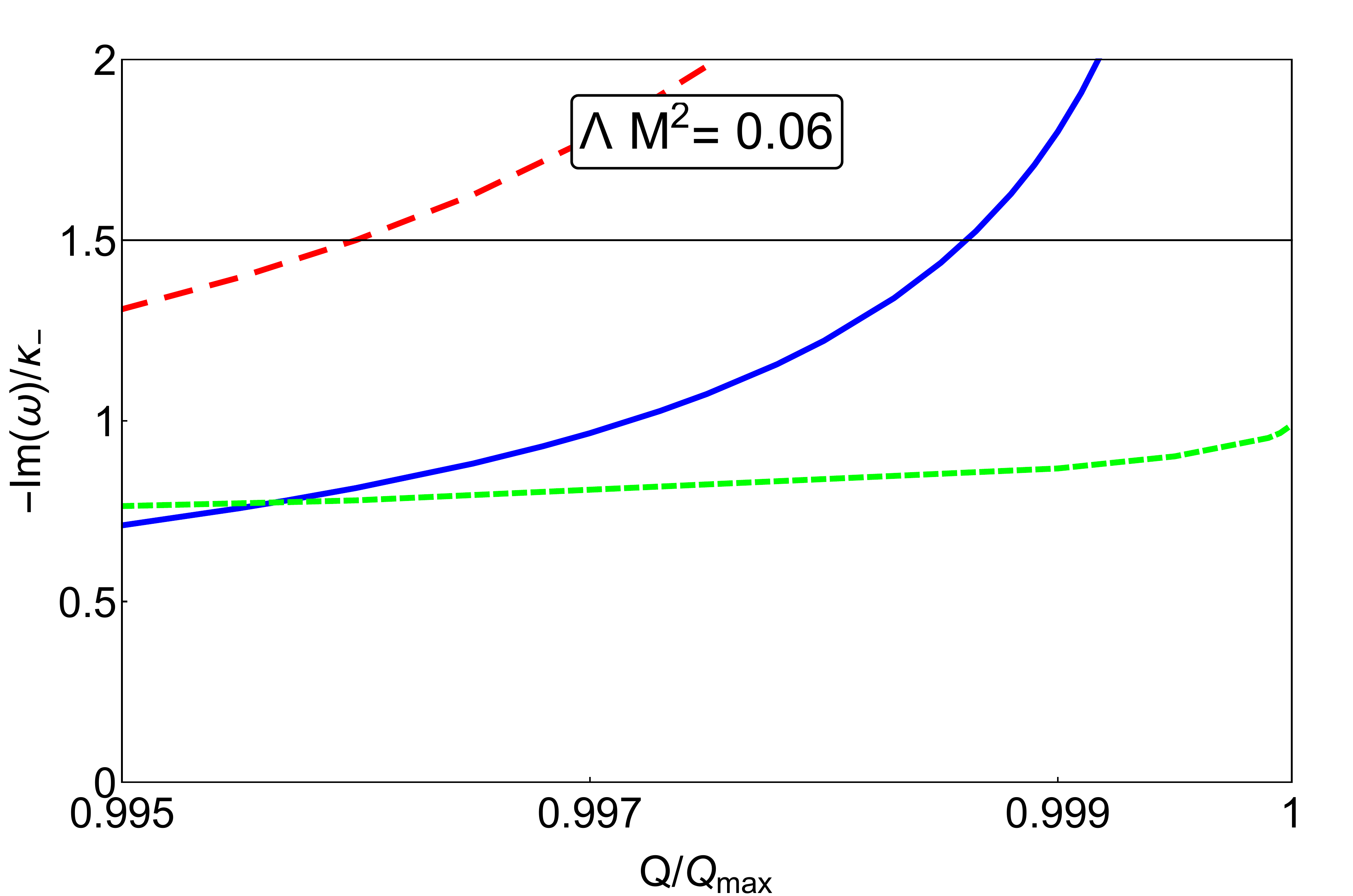}
\includegraphics[scale=0.2]{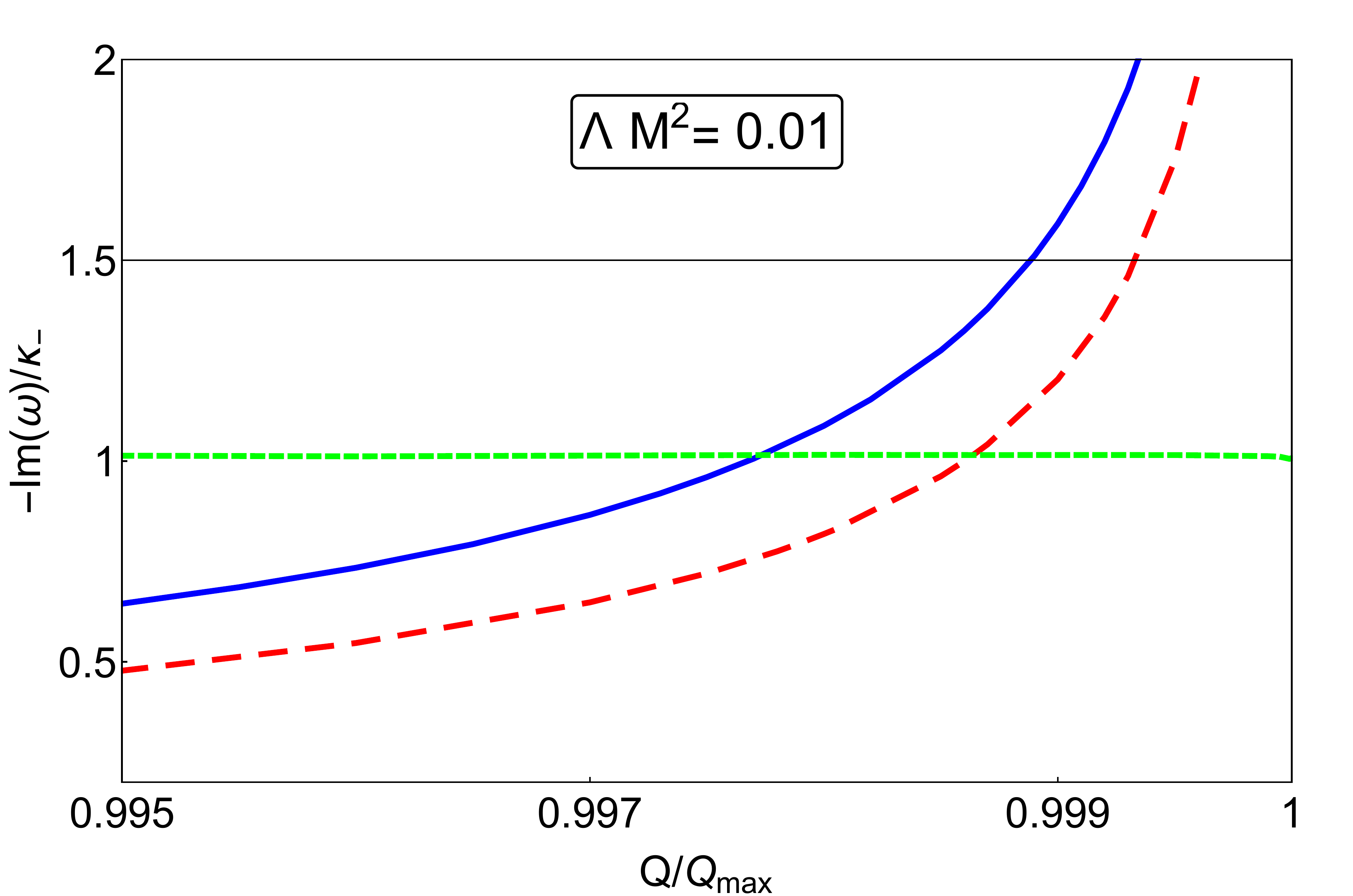}
\includegraphics[scale=0.2]{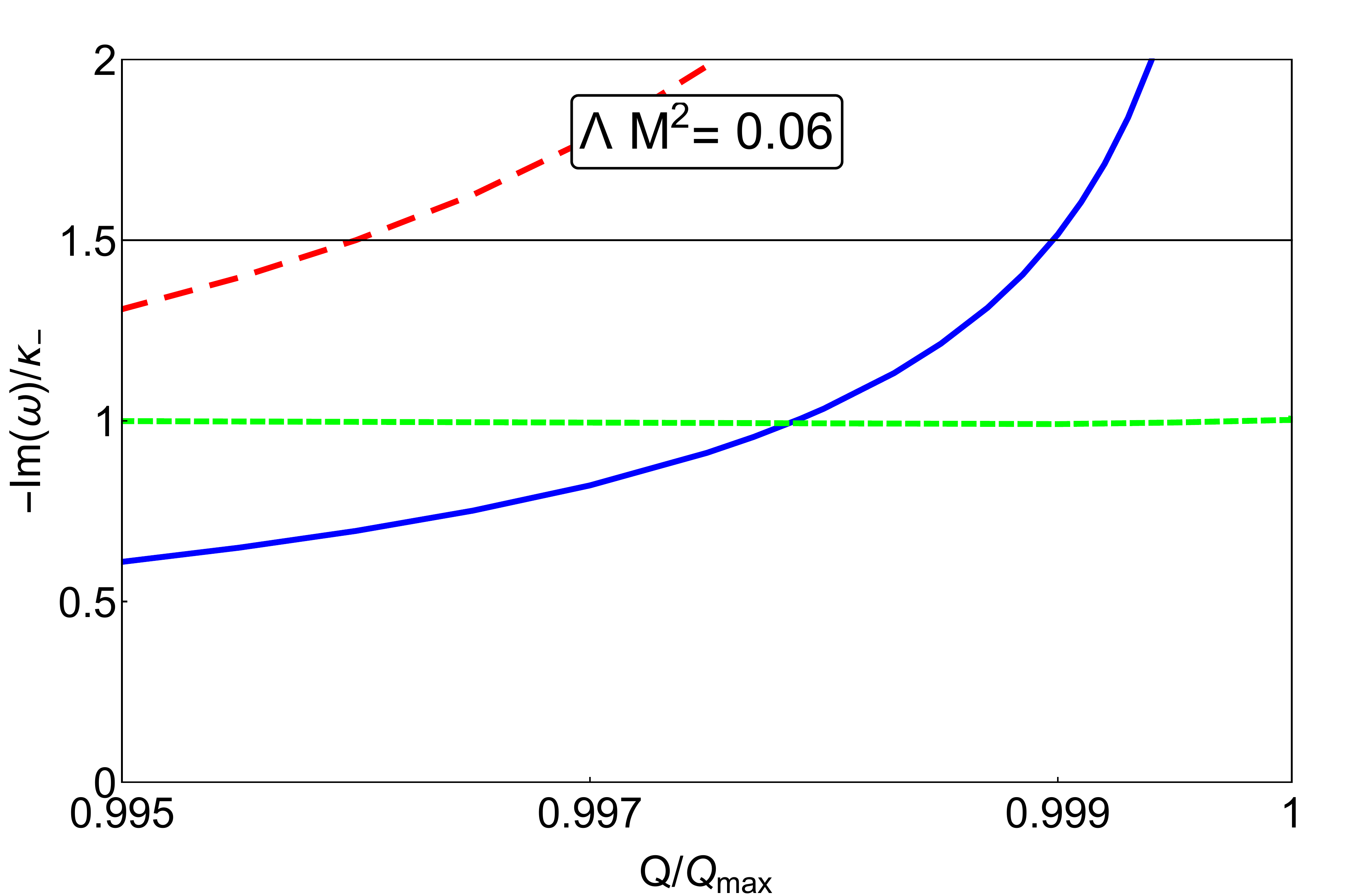}
\caption{Imaginary parts (divided by the surface gravity $\kappa_-$ of the CH) of the dominant families of modes of a non-minimally coupled scalar field propagating on a fixed near-extremal RNdS background. The top panels show the behavior of the modes for $\eta=-0.5$, while the bottom ones show the behavior of the modes for $\eta=0.5$. The dominant (approximated with $l=10$) PS modes are depicted with blue lines, the dominant ($l=1$) dS modes with dashed red lines and the dominant ($l=0$) NE modes with dotted green lines.}
\label{QNMs}
\end{figure}

{\bf The photon sphere family:} The photon sphere (PS) QNMs, are represented by damped oscillations whose decay rate, $\omega_I$, is directly connected to the instability timescale of null geodesics at the photon sphere. For this family of modes, higher angular momentum represents smaller decay rates such that, in the limit of interest, the most representative modes are those for which $l \rightarrow \infty$. By inspection \cite{Cardoso:2017soq,Destounis:2018qnb,Liu:2019lon}, we find that $l=10$ provides a good approximation for the imaginary part of the dominant mode (relative to $\beta$).

{\bf The de Sitter family:} The de Sitter (dS) family of modes are related to the accelerated expansion of the Universe which, in turn, is related to the surface gravity of the cosmological horizon of pure dS space \cite{Brill:1993tw,Rendall:2003ks}. They correspond to purely imaginary modes which can be very well approximated by the pure dS QNMs  \cite{Du:2004jt,LopezOrtega:2006my,VasydS}
\begin{align}
\label{dS1}
\omega_{n=0,\text{pure dS}}/\kappa_c^\text{dS}&=-i l,\\
\label{dS2}
\omega_{n\neq 0, \text{pure dS}}/\kappa_c^\text{dS}&=-i (l+n+1),
\end{align}
where $\kappa_c^\text{dS}=\sqrt{\Lambda/3}$ is the surface gravity of the cosmological horizon of pure dS space and $n$ is the overtone number. The dS family of modes has a surprisingly negligible dependence on the BH charge and seems to be well described only by $\kappa_c^\text{dS}$. Such expressions are exactly the same for the Horndeski action studied here, and are only modified in the massive scalar field case (see e.g. Eq. (22) in \cite{Fontana:2018fof}). The dominant mode is obtained for $n=0$, $l=1$ and is almost identical to \eqref{dS1}, while higher overtones have increasingly larger deformations to \eqref{dS2}.

{\bf The near-extremal family:} The near-extremal (NE) family of modes arises and dominates the dynamics of the ringdown waveform at late times, for very high values of the BH charge ($r_-\sim r_+$). For the RNdS spacetime in GR, the modes of this family approach
\begin{equation}
\label{NE}
\omega_\text{NE}=-i(l+n+1)\kappa_{-}=-i(l+n+1)\kappa_{+},
\end{equation}
where $\kappa_-$ and $\kappa_+$ are the surface gravity of the Cauchy and event horizon, respectively, in the RNdS spacetime. The dominant mode is obtained for $n=l=0$. Higher angular numbers $l$ admit larger (in absolute value) imaginary parts, thus being irrelevant to our discussion.

In the particular Horndeski theory, our numerics indicate that \eqref{NE} is highly affected by the introduction of $\eta$. A negative $\eta$ makes the NE modes decay slower, while a positive $\eta$ makes the NE modes decay faster. An analytic expression for this family is, unfortunately, lacking.
\begin{figure}[t]
\centering
\includegraphics[scale=0.2]{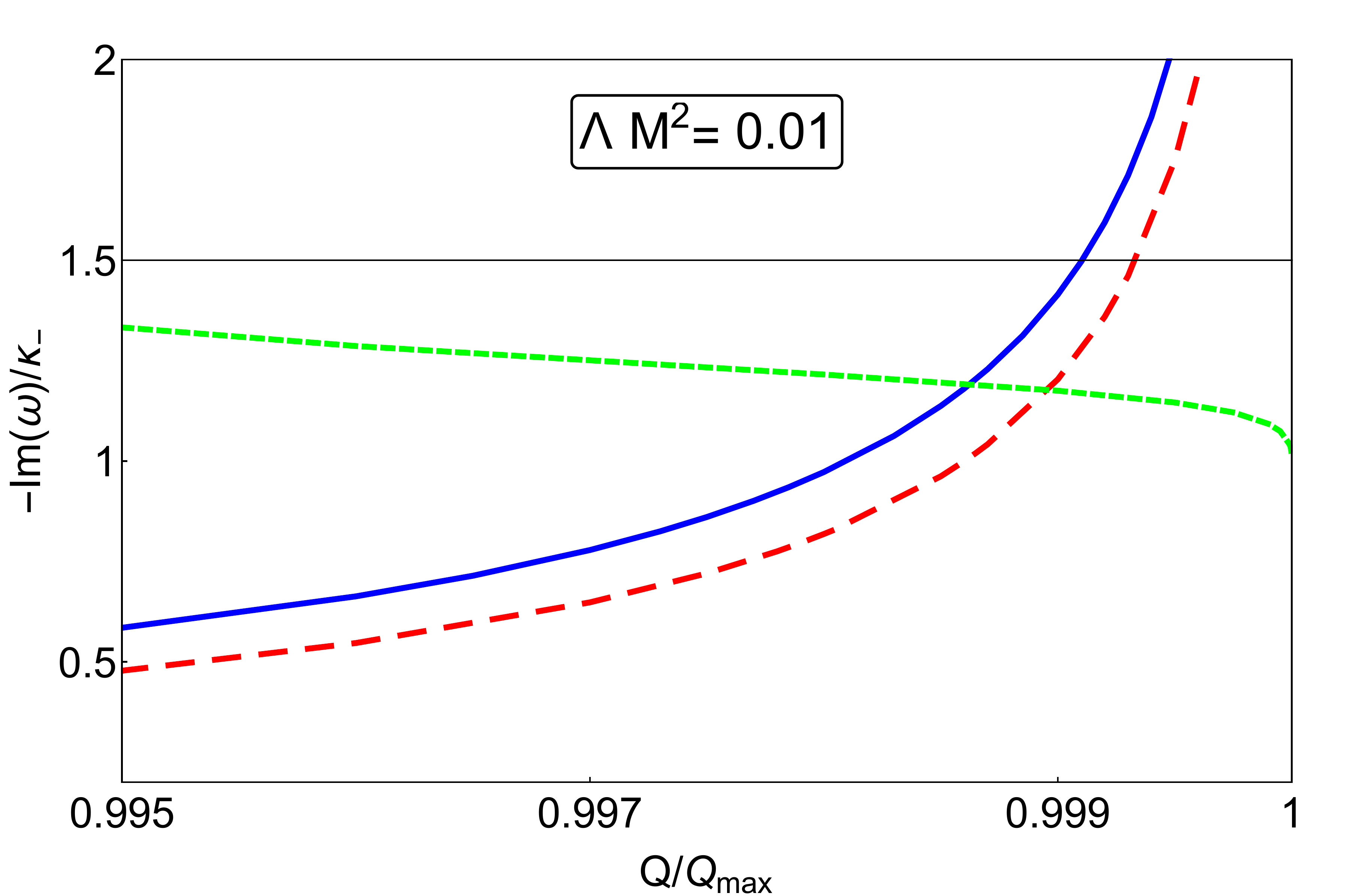}
\includegraphics[scale=0.2]{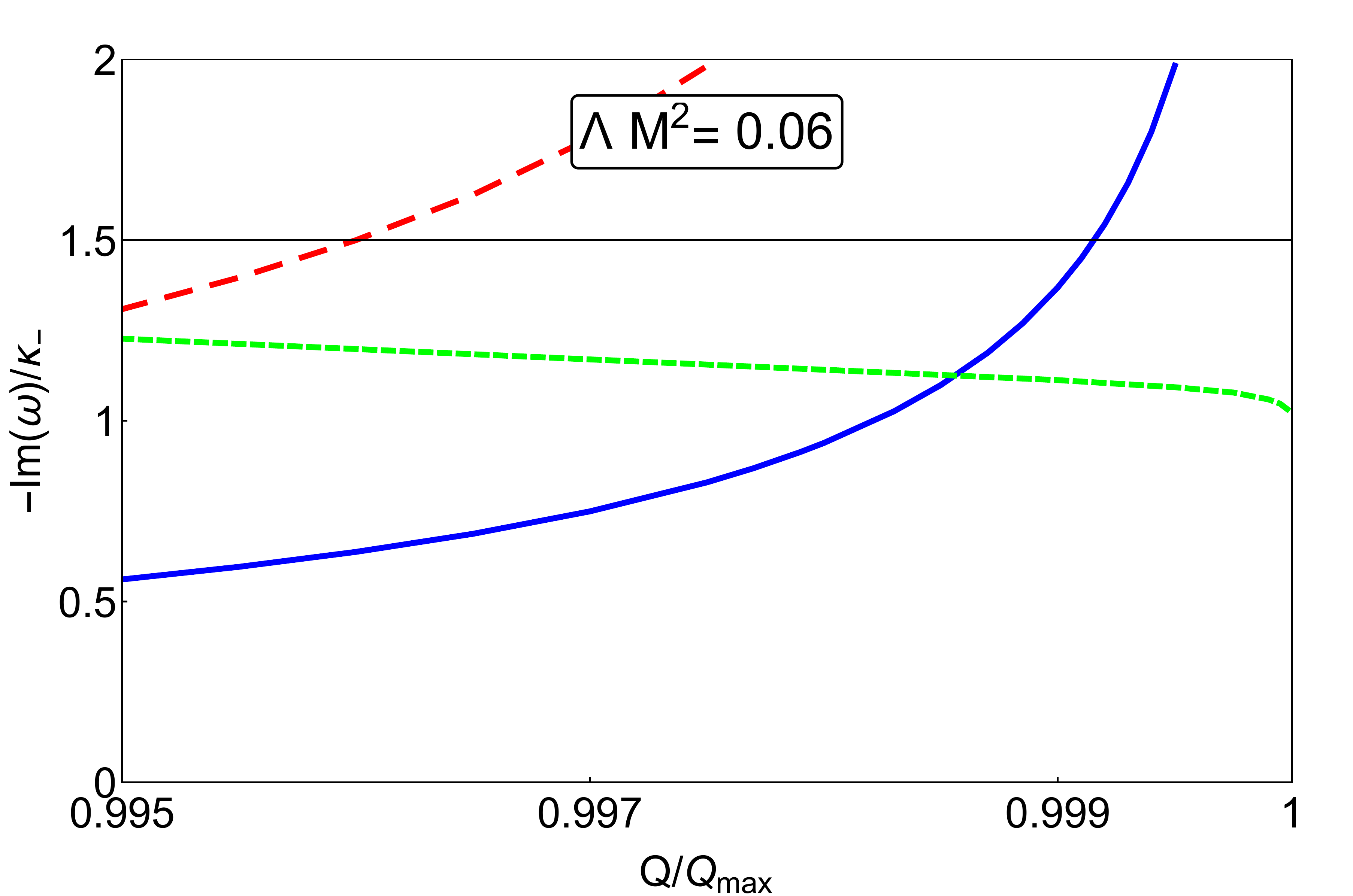}
\includegraphics[scale=0.2]{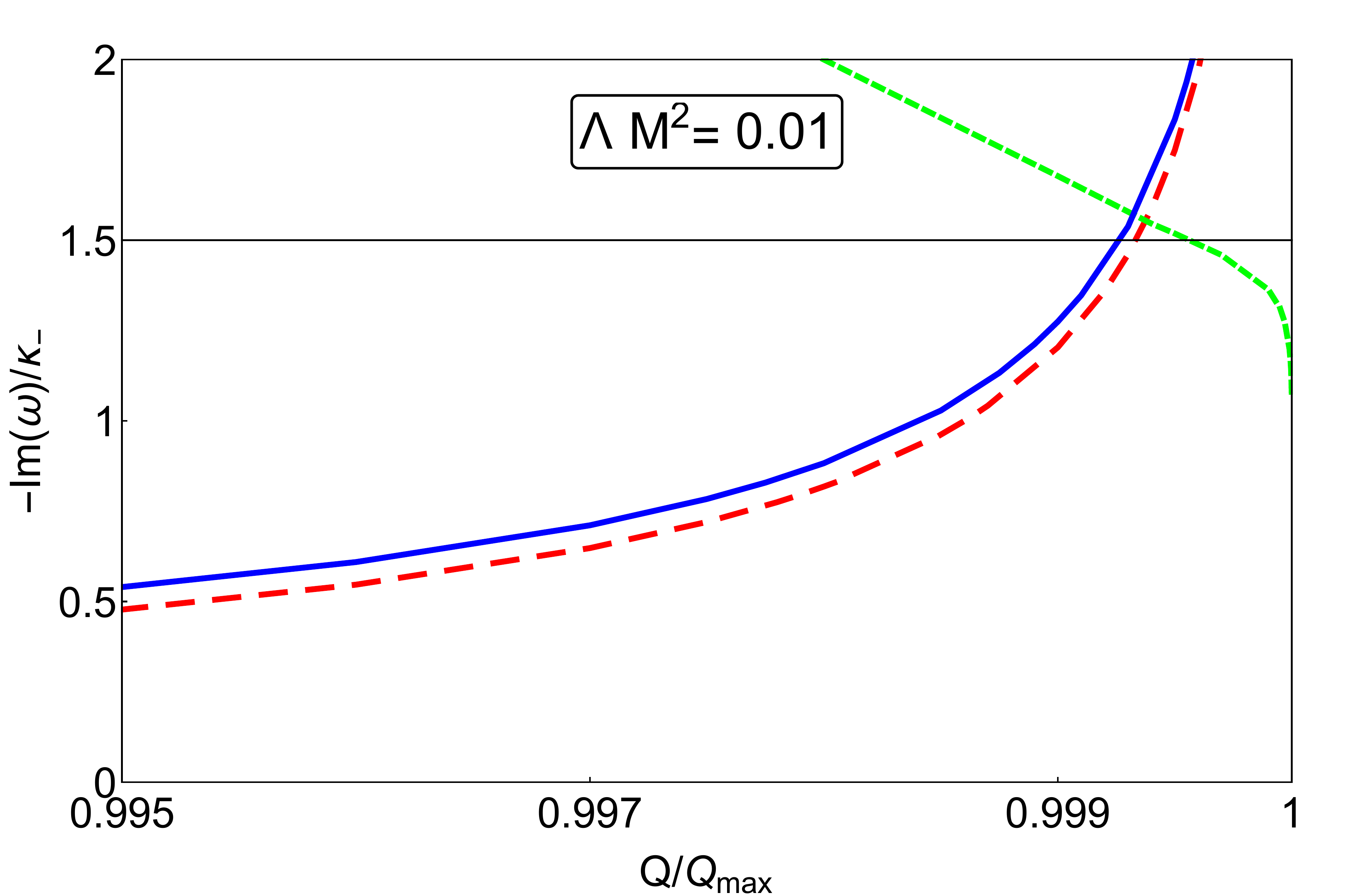}
\includegraphics[scale=0.2]{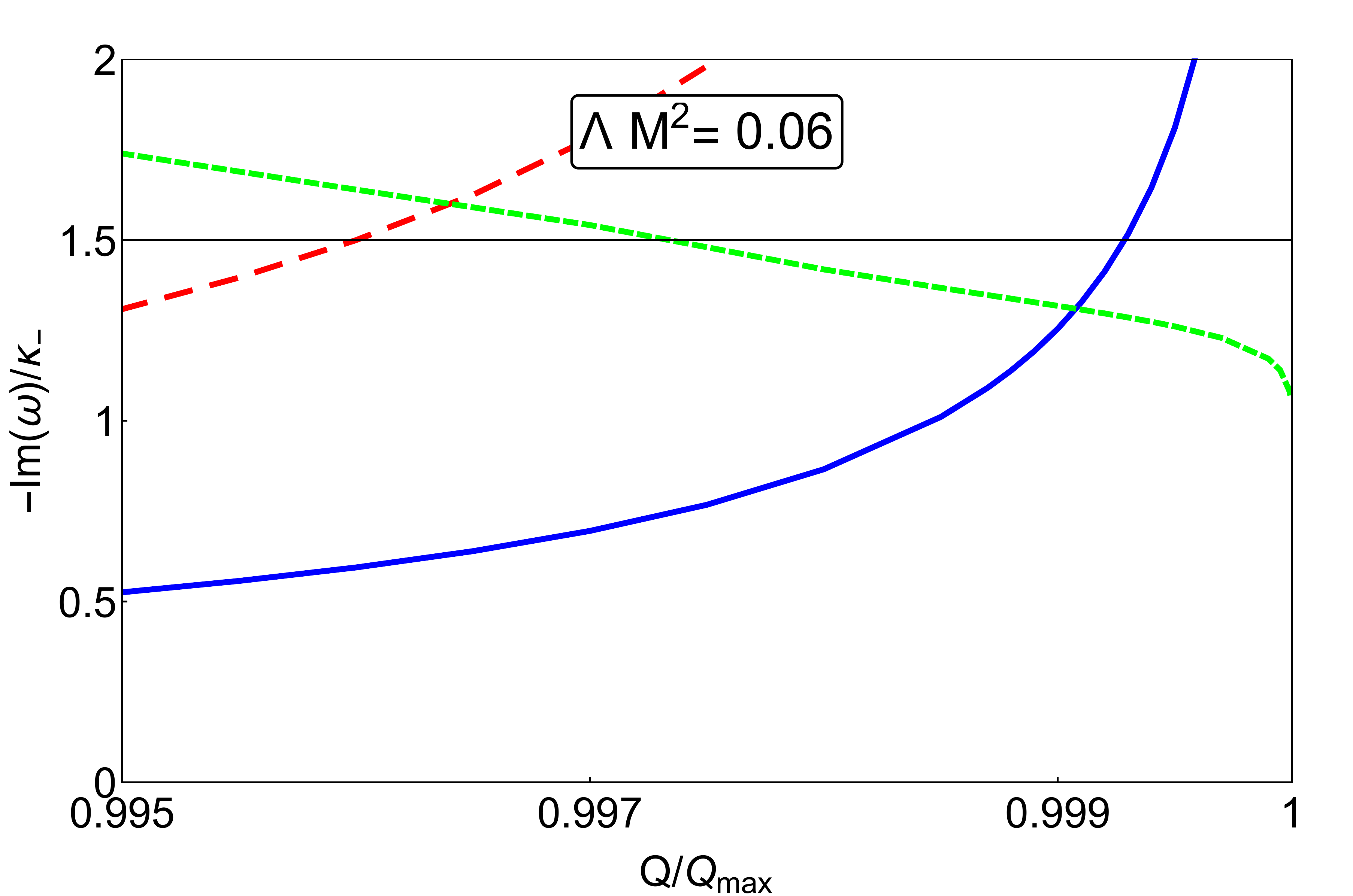}
\caption{Imaginary parts (divided by the surface gravity  $\kappa_-$ of the CH) of the dominant families of modes of a non-minimally coupled scalar field with $\eta=0.8$ (top panel) and $\eta=0.975$ (bottom panel), propagating on a fixed near-extremal RNdS background. The dominant (approximated with $l=10$) PS modes are depicted with blue lines, the dominant ($l=1$) dS modes with dashed red lines and the dominant ($l=0$) NE modes with dotted green lines.}
\label{eta08}
\end{figure}

In Appendix \ref{appA}, we show, numerically, that the late-time behavior of scalar perturbations propagating on RNdS BHs in the particular Horndeski theory, is governed by the dominant QNMs. Furthermore, perturbations always decay exponentially and no instabilities are found for the couplings considered ($|\eta|<1$). As previously discussed, $\eta$ captures the deviation from classical GR and large enough couplings $\eta$ lead to instabilities of the BH exterior \cite{Fontana:2018fof}. The onset of these linear instabilities designates the point beyond which the BH itself senses the modification of gravity, and this either heads to the scalarization of the BH and the formation of a new stable object or to the dispersion of all matter. We will not be interested in $|\eta|>1$, since the discussion of SCC would be irrelevant in regions of instabilities of the RNdS BH in study.

In Fig. \ref{QNMs} and Fig. \ref{eta08}, we display the interplay between the three families of modes for several values of $\Lambda$ and $\eta$. The PS family, depicted with blue color, seems to decay faster (resp. slower) for negative (resp. positive) $\eta$. This can be explained with the following argument: the coupling $\eta$ introduces a new scale in the theory. If $\eta$ is negative, then it acts as a friction term absorbing energy from the kinetic energy of the scalar field \cite{Papantonopoulos:2019eff}. This means that the scattered wave has less kinetic energy to maintain its evolution, so it decays faster. On the contrary, if $\eta$ is positive this effect does not occur and the scattered wave maintains its energy.\footnote{This can also be understood as an effect of the geometry. For example, in a cosmological model the curvature effects are strong during inflation absorbing energy, and it was found in \cite{Sushkov:2009hk} that a negative $\eta$ leads to a fast collapse of the inflaton field to the initial singularity, while, after inflation, curvature is small and  a positive $\eta$ leads the universe to exit the quasi-de Sitter phase.} The PS family seems to dominate the dynamics of the ringdown at late times for large enough $\Lambda$ (see right panels of Fig. \ref{QNMs}). In these regions of the parameter space, we expect $\beta$ to be defined by the PS modes (until the NE family takes over).
\begin{figure}[t]
\centering
\includegraphics[scale=0.2]{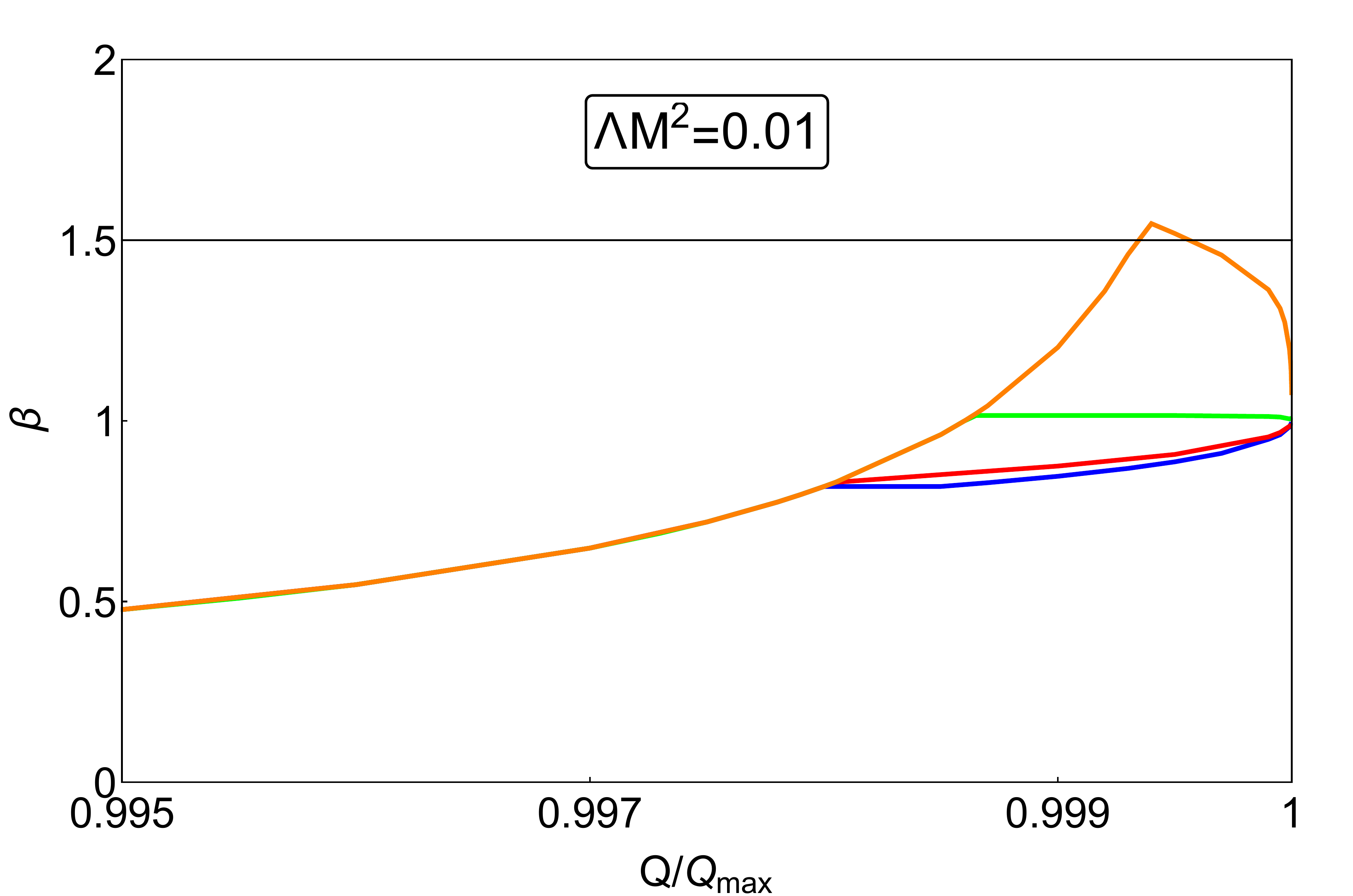}
\includegraphics[scale=0.2]{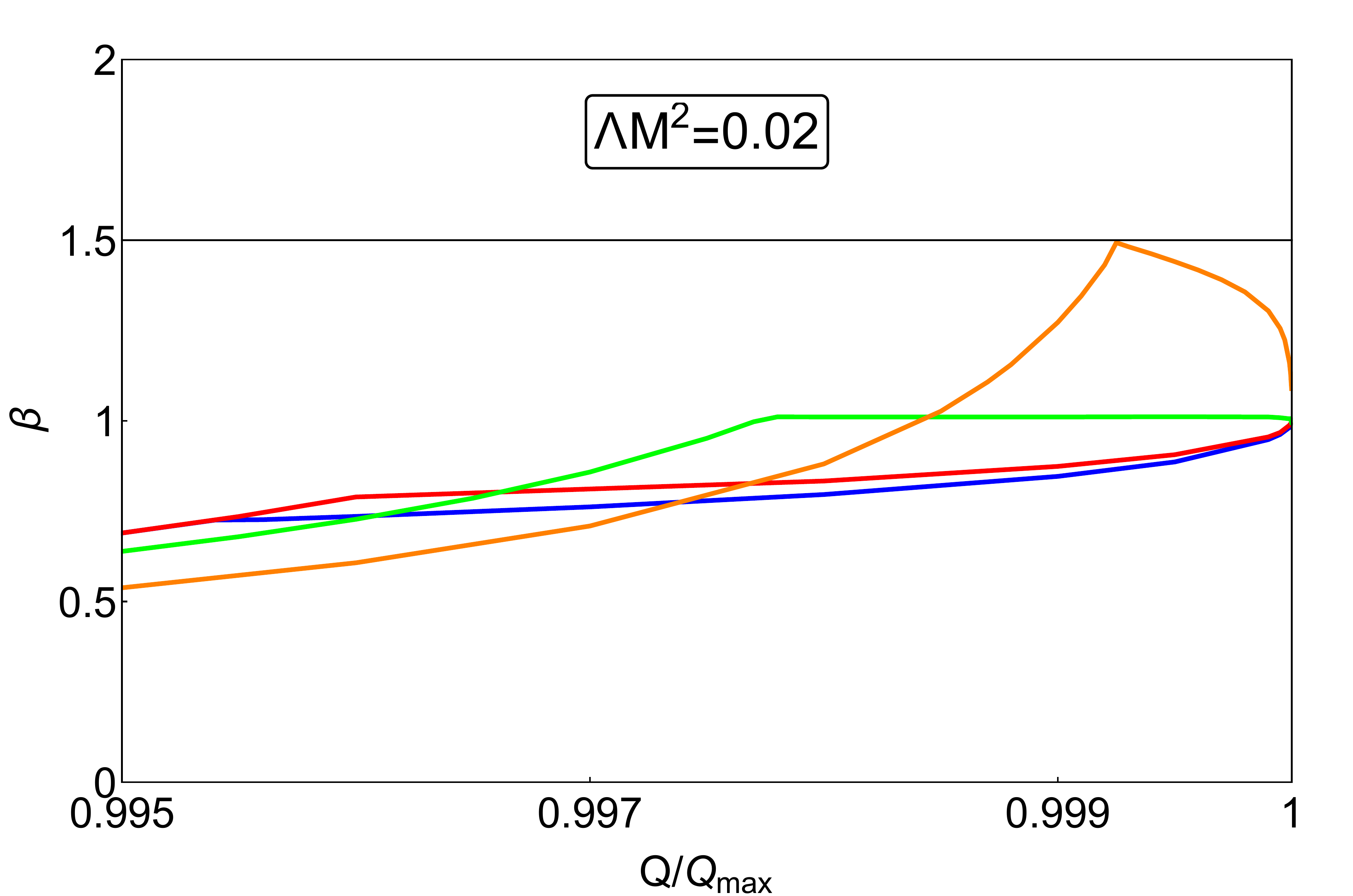}
\includegraphics[scale=0.2]{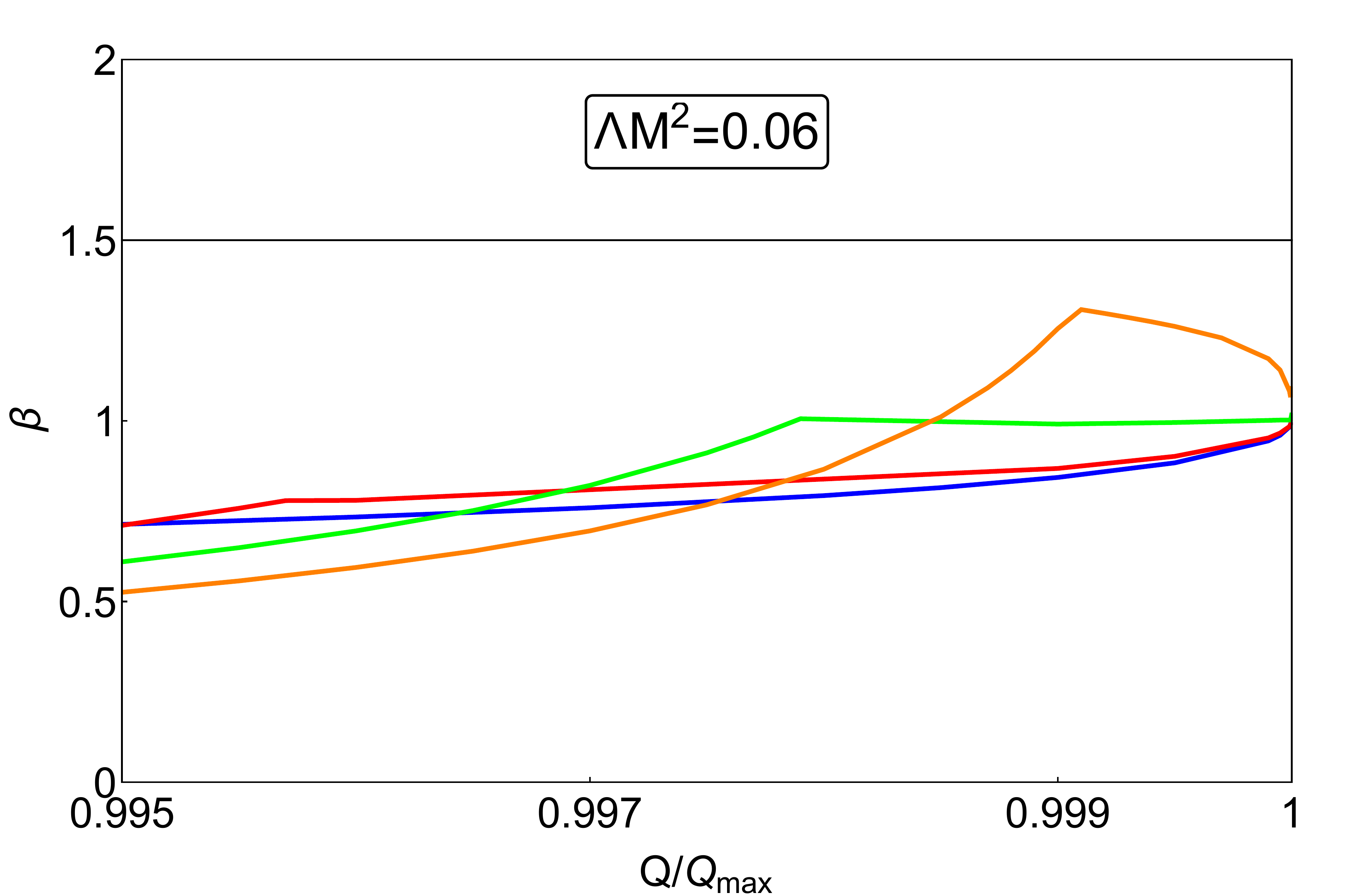}
\includegraphics[scale=0.2]{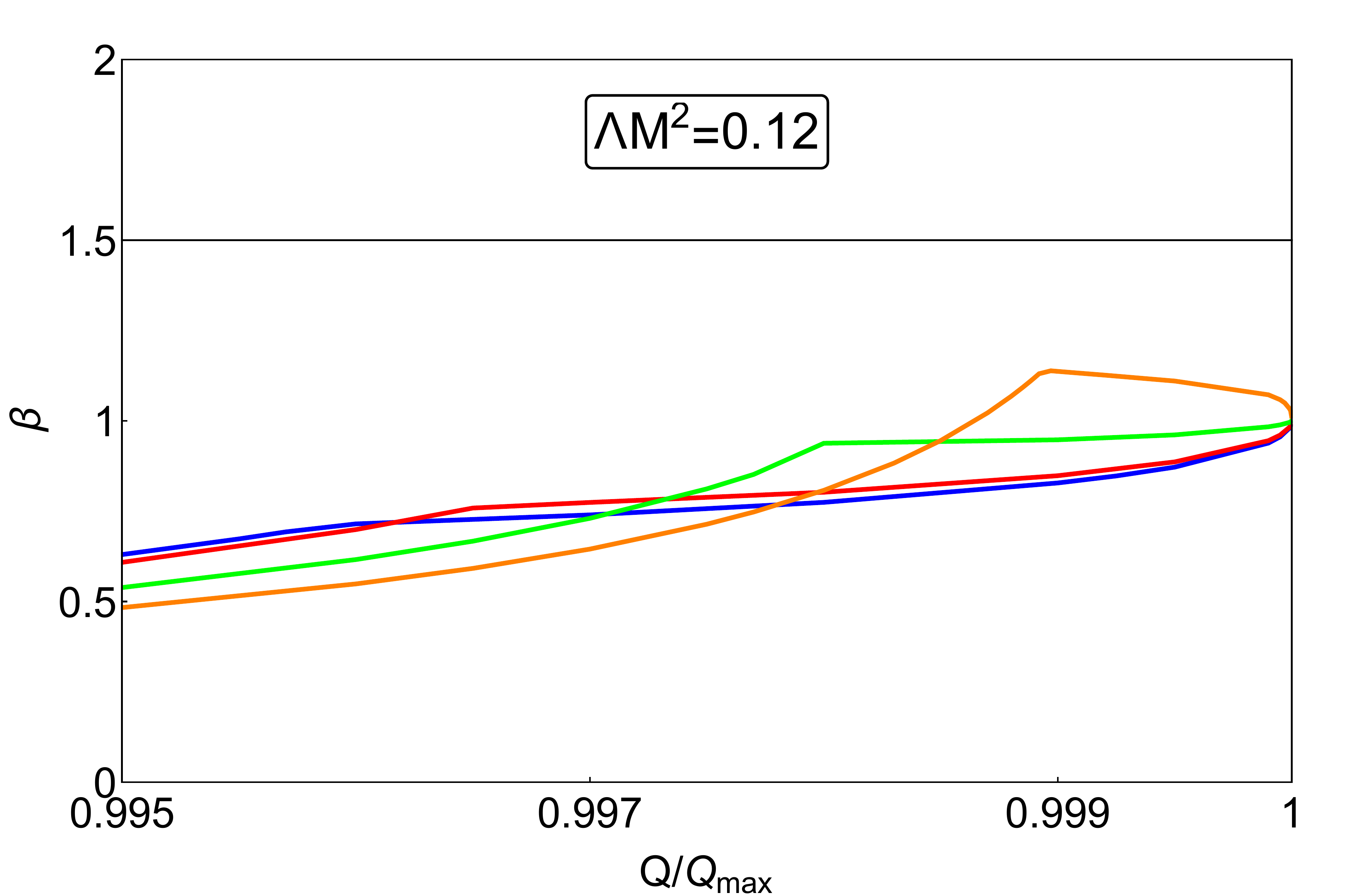}
\caption{The parameter $\beta$ calculated from the dominant QNMs of a non-minimally coupled scalar field propagating on a fixed near-extremal RNdS background. The black horizontal line denotes $\beta=3/2$. The values of $\eta$ shown are $-0.975$ (blue), $-0.5$ (red), $0.5$ (green) and $0.975$ (orange).}
\label{beta}
\end{figure}

As expected, the dS family remains unaffected under the introduction of $\eta$. Moreover, it is dominant for small $\Lambda$ as expected from \cite{Cardoso:2017soq,Destounis:2018qnb} (see left panels of Fig. \ref{QNMs}). Hence, for small $\Lambda$, we expect $\beta$ to be defined by the dS modes (until the NE family takes over).

Finally, the NE family, indeed, dominates close to extremality and decays faster (resp. slower) for $\eta>0$ (resp. $\eta<0$). In Figs. \ref{QNMs}, \ref{eta08}, we can always find a region in the parameter space where the NE family is the one with the smallest imaginary part contribution and, therefore, $\beta$ will be defined by the NE modes in these regions. We can, therefore, realize that a possible violation of SCC might occur for large enough $\eta>0$, since $-\text{Im}(\omega)/\kappa_-\rightarrow 1$ from above (see Fig. \ref{eta08}).

\section{The fate of Strong Cosmic Censorship}
\begin{figure}[t]
\centering
\includegraphics[scale=0.4]{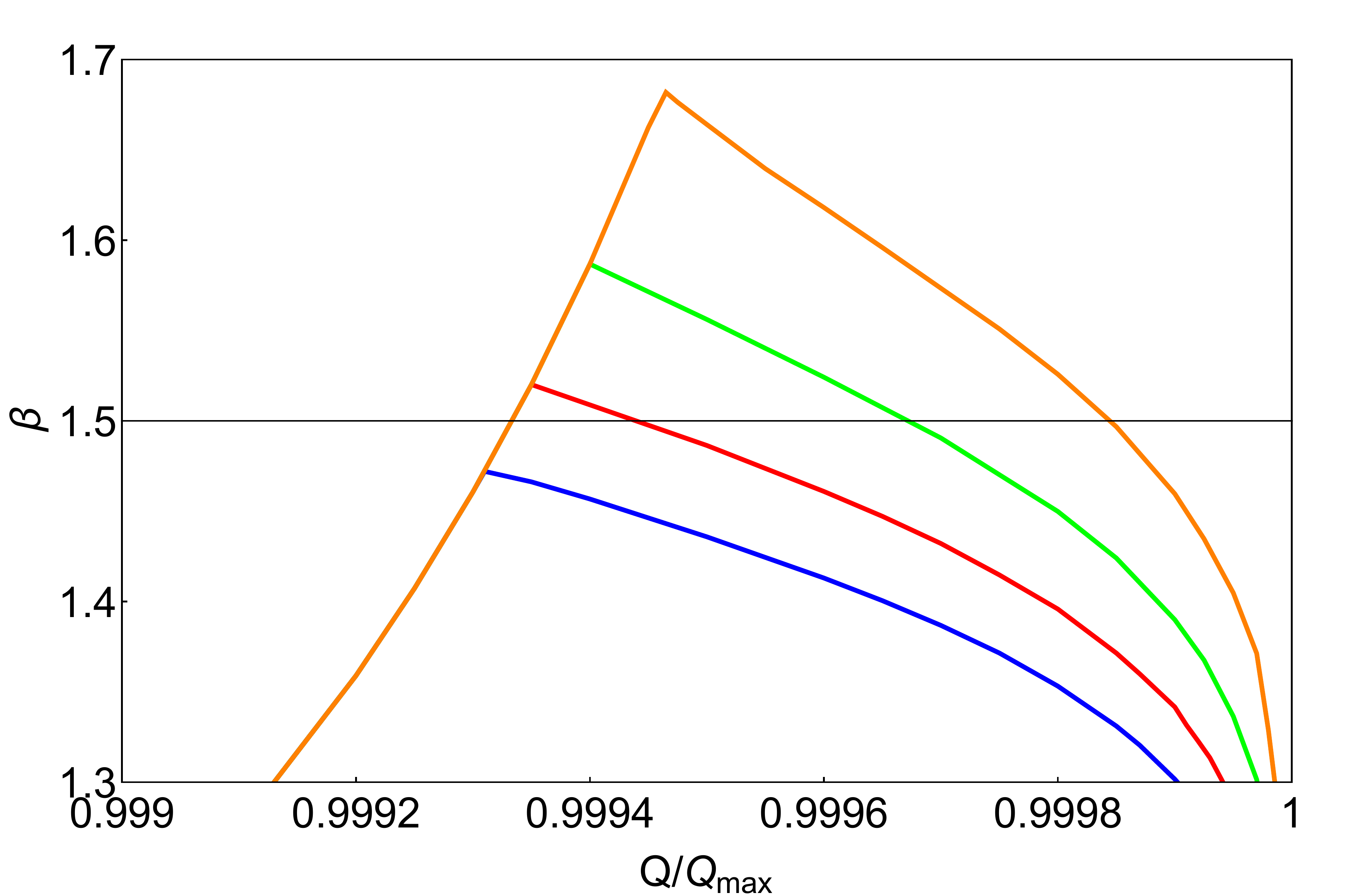}
\caption{The parameter $\beta$ calculated from the dominant QNMs of a non-minimally coupled scalar field propagating on a fixed near-extremal RNdS background with $\Lambda M^2=0.01$. The black horizontal line denotes $\beta=3/2$. The values of $\eta$ shown are $0.96$ (blue), $0.97$ (red), $0.98$ (green) and $0.99$ (orange).}
\label{betamax}
\end{figure}
The interplay between the families of modes with respect to the non-minimal coupling $\eta$ will decide the fate of SCC. Recall that, if $\beta>3/2$, weak solution of the field equation can exist at the CH and can be extended beyond it, thus violating the SCC conjecture.

To calculate $\beta$ we always choose the smallest, non-zero, contribution of imaginary parts from all the families of modes. In Fig. \ref{beta}, we plot $\beta$ for various choices of $\eta$ and $\Lambda$ for near-extremally charged RNdS BHs. The sharp transitions on the figures designate the points, in the parameter space, where the NE family begins dominating the dynamics against either the PS or dS family, as previously discussed.
Different colors in the plots designate different choices of $\eta$.

We observe that, as $\eta$ increases from negative to positive, the interplay of the different families of QNMs gives rise to regions with larger or smaller $\beta$, depending on the cosmological constant, until the point where the NE modes dominate and terminate $\beta$ at unity, for extremal BH charge. In the region where the NE family dominates, we clearly see a pattern which indicates that the increment of $\eta$ leads to the increment of $\beta$. More precisely, when $\eta<0$ we observe that $\beta<1$ always, indicating that in such regions SCC is respected in the particular Horndeski theory. When $\eta>0$, $\beta$ is not bounded by an absolute threshold but depends on the choice of $\Lambda$.

On the contrary, the increment of $\Lambda$ has the ability to decrease $\beta$ near extremality. The effects of increasing $\Lambda$ and $\eta$ seem to counterbalance each other, for most of the volume of the subextremal parameter space, which leads to $\beta<3/2$. For all these regions, SCC should be respected.

For small $\Lambda$ and large enough $\eta$, $\beta$ can exceed $3/2$, according to our numerics, thus putting the validity of SCC into question. On the top left panel of Fig. \ref{beta} (see also bottom left plot in Fig. \ref{eta08}), for $\Lambda M^2=0.01$ and $\eta=0.975$, we do find a very small region where $\beta>3/2$. In Fig. \ref{betamax}, we zoom into this region and increase $\eta$ smoothly. We demonstrate that as $\eta\rightarrow1$, the violation gap increases. We expect a similar behavior for even smaller cosmological constants. Hence, in this region of the parameter space, where $\beta>3/2$, the SCC conjecture is violated in the particular Horndeski theory.

We can safely assume that as $\eta$ increases even more, the violation gap will be enlarged. Such deviation from GR, though, leads to discontinuities and instabilities, at the linear level, as previously discussed (see \cite{Fontana:2018fof}). Therefore, more increment of $\eta$ beyond unity is futile, since for RNdS BHs with an unstable exterior, the discussion of the validity of SCC is redundant.

\section{Conclusions}

The modern formulation of strong cosmic censorship, proposed by Christodoulou, states that appropriately chosen initial data should be future inextendible beyond the Cauchy horizon, as a suitable metric with square-integrable Christoffel symbols. The conjecture includes spacetimes which possess a Cauchy horizon, like the ones describing charged and rotating black holes.

As recently shown in \cite{Cardoso:2017soq, Luna:2018jfk}, an electrically charged black hole in de Sitter spacetime with a scalar field poses a serious threat to the validity of strong cosmic censorship. If a near-extremally charged black hole could exist in our Universe, then macroscopic observers would, in principle, reach a region deep inside the black hole interior where the deterministic nature of physical laws breaks down and the observers would live a highly unpredictable future. The passage beyond the Cauchy horizon would be smooth enough for the observers to not be destroyed \cite{Ori:1991zz}, while the field equations would still make sense \cite{Klainerman:2012wt}.

Luckily for all of us, there are no near-extremal electrically charged black holes, as far as we know. In fact, most black holes seem to be almost neutral due to various dissipation processes. From the present physical point of view, it would be interesting, then, to study the fate of strong cosmic censorship in rapidly rotating black holes, which seem to exist in nature \cite{2011ApJ...736..103B,FastSpin}. An answer, given recently in \cite{Dias:2018ynt}, appears to be positive; rapidly rotating black holes in de Sitter spacetime seem to respect strong cosmic censorship. Although this might be seen as the end of scenarios of realistic astrophysical black holes violating determinism, if we consider strong cosmic censorship as a mathematical way of testing classical General Relativity and its limits, then the violation of strong cosmic censorship in Einstein-Maxwell theory is not to be taken lightly.

One, then, would wonder if slight modifications to General Relativity could heal such a fragility and restore predictability even for spherically symmetric spacetimes. In this study, we have considered a non-minimal coupling between the Einstein tensor and a probe scalar field, which propagates on a fixed Reissner-Nordstr\"om-de Sitter background. Such a higher-order derivative coupling theory belongs to the Horndeski scalar-tensor class. The theory modifies the kinematic properties of the scalar field and, more importantly, makes the regularity requirements, for the existence of weak solutions of the field equations at the Cauchy horizon, even stronger. As a matter of fact, we have shown that the parameter $\beta$ that decides the fate of strong cosmic censorship, in this theory, requires $\beta>3/2$ for violation to occur, in contrast to previously reported studies where the corresponding condition was $\beta>1/2$. This novel requirement makes the formation of a singularity at the Cauchy horizon more likely, in a sense, and therefore, a stable enough Cauchy horizon seems harder to obtain under scalar perturbations.

The setup seems very promising since, for most of the volume of the parameter space, strong cosmic censorship appears to be respected. However, our study indicates that we can still find regions in the sub-extremal parameter space of ``small'' near-extremal Reissner-Nordstr\"om-de Sitter black holes, with positive non-minimal couplings, where the conjecture is not respected. Why spherical symmetry is unfavored, even in this theory, remains unknown.

\section*{Acknowledgements}
The authors are grateful to Jo\~ao Costa for helpful discussions. KD acknowledges financial support provided under the European Union's H2020 ERC Consolidator Grant "Matter and strong-field gravity: New frontiers in Einstein's theory" grant agreement no. MaGRaTh--646597. KD also acknowledges networking support by the GWverse COST Action CA16104, "Black holes, gravitational waves and fundamental physics". FCM thanks hospitality of the AEI in Golm, FCT project PTDC/MAT-ANA/1275/2014, CAMGSD, IST, Univ. Lisboa, through FCT project UID/MAT/04459/2013, CMAT, Univ. Minho, through FCT project Est-OE/MAT/UI0013/2014 and FEDER Funds COMPETE.
\appendix
\section{The late-time behavior of scalar perturbations propagating on Reissner-Nordstr\"om-de Sitter black holes in Horndeski theory}\label{appA}

The late-time behavior of perturbations is a crucial ingredient for the study of the CH instability and the fate of SCC. It is already proven \cite{Hintz:2016gwb,Hintz:2016jak} that perturbations of Kerr-de Sitter and Kerr-Newman-de Sitter BHs decay at late times, following an exponential Price law. These results account for the linear and non-linear stability of this class of solutions. Moreover, the decay of perturbations is governed by the dominant QNMs of the spacetime. In particular, it has rigorously been shown that, for some $\phi_0\in \mathbb{C}$,
\begin{equation}
\label{exponential tail}
|\phi-\phi_0|\leq C e^{-\alpha t},
\end{equation}
with $\phi_0$ a constant shift to the scalar field configuration representing the ``zero-mode'' of the dS family, and $\alpha=-\text{Im}(\omega)$, the spectral gap, i.e. the imaginary part of the lowest-lying/dominant, non-zero QNM, $\omega$.

In this section, we shall provide strong numerical evidence which indicates that \eqref{exponential tail} still holds when the scalar field, which propagates on a fixed RNdS background, is coupled to the Einstein tensor. To do so, we depict four different cases, where various families dominate the late-time behavior of the perturbation.

\begin{figure}[t]
\centering
\includegraphics[scale=0.205]{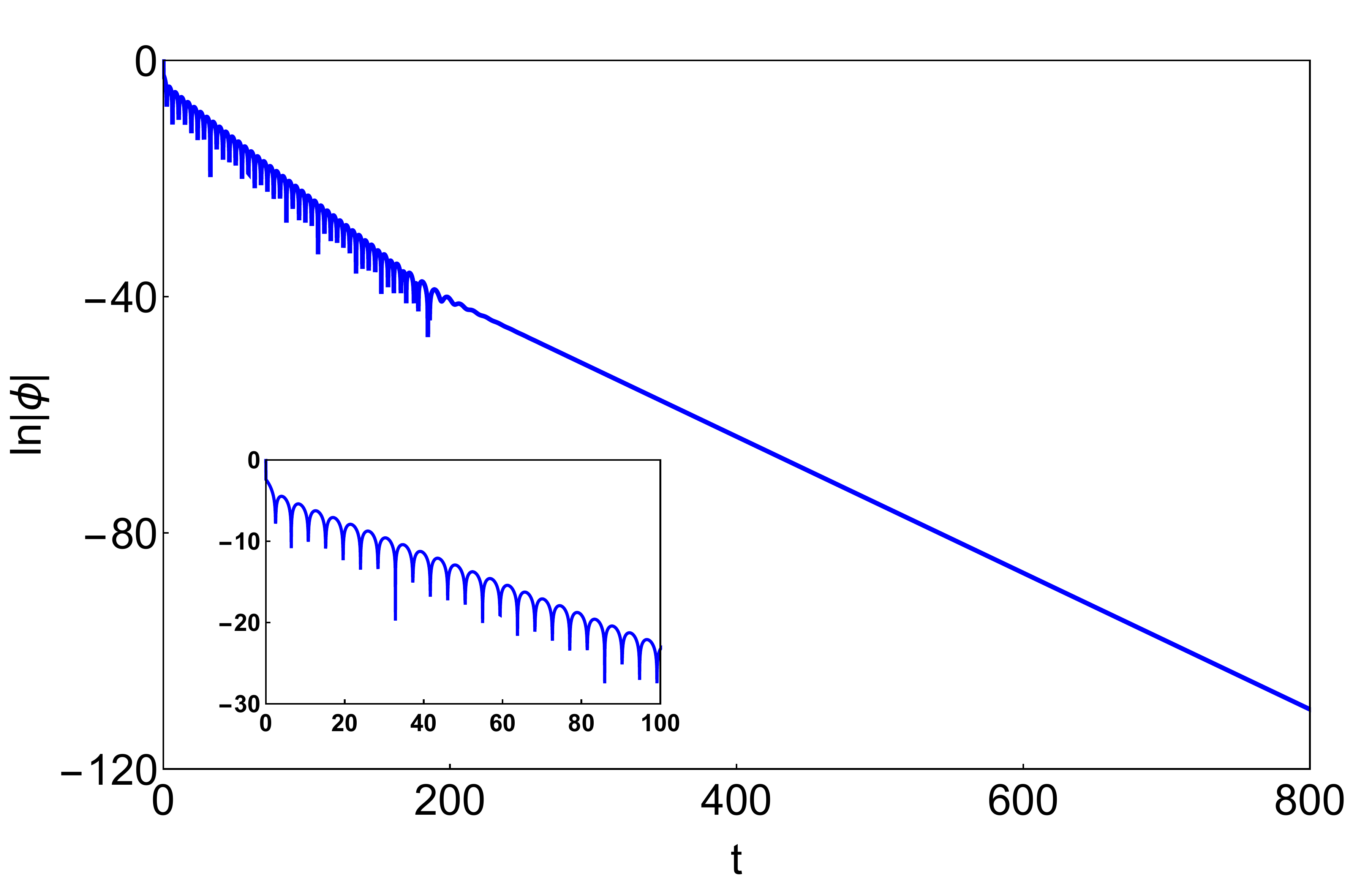}\hskip -1ex
\includegraphics[scale=0.2]{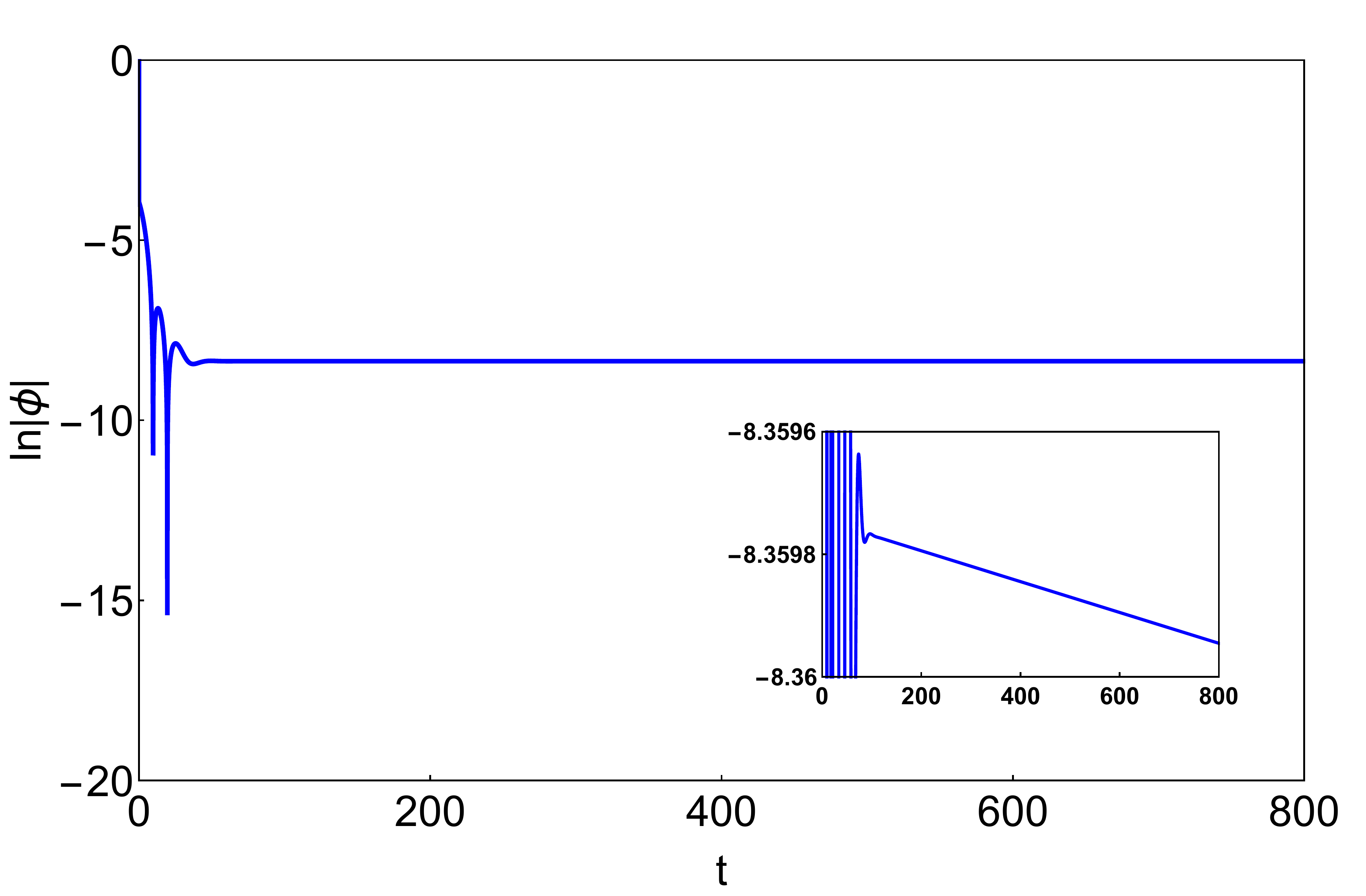}
\includegraphics[scale=0.205]{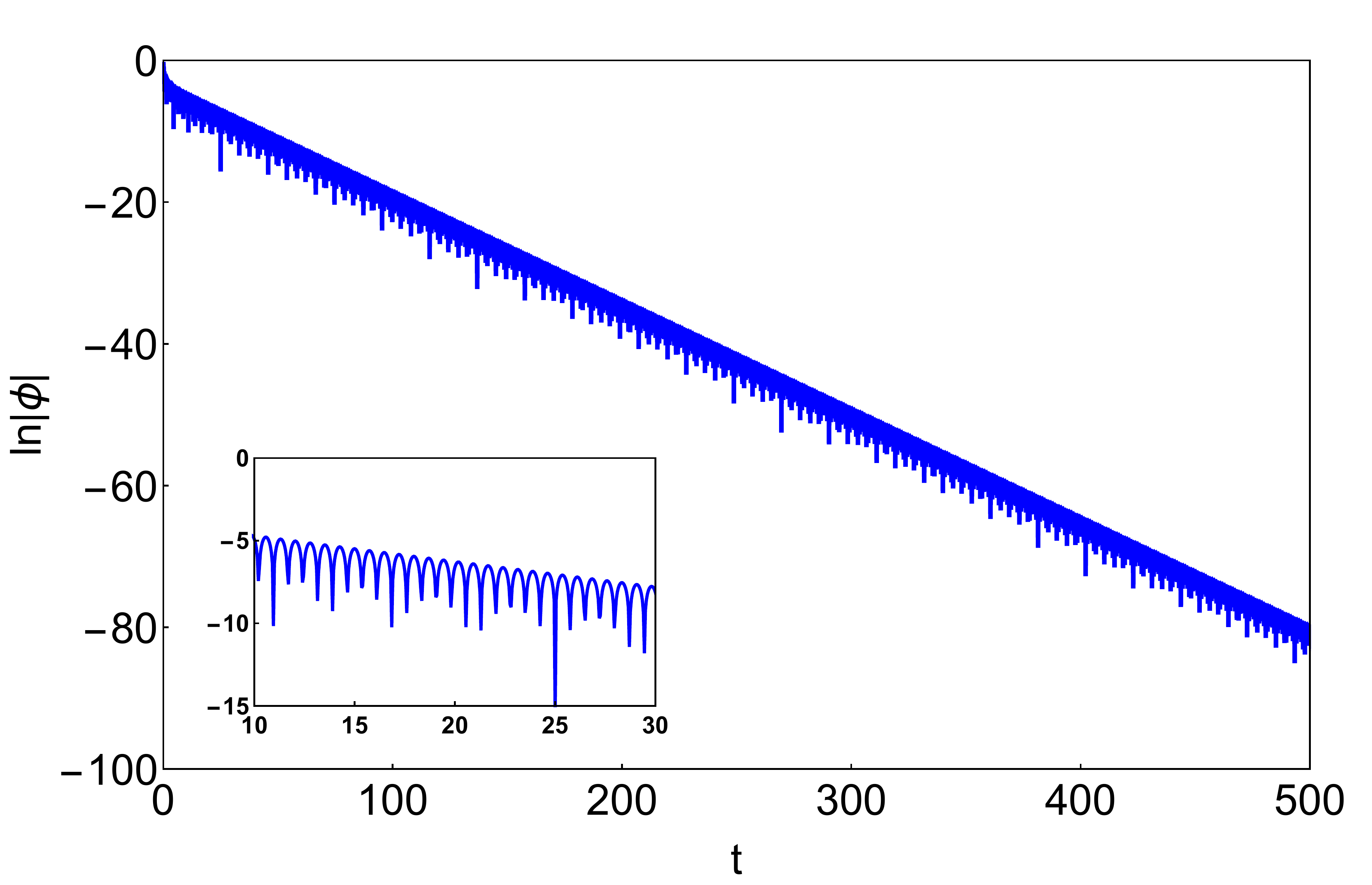}\hskip -1ex
\includegraphics[scale=0.2]{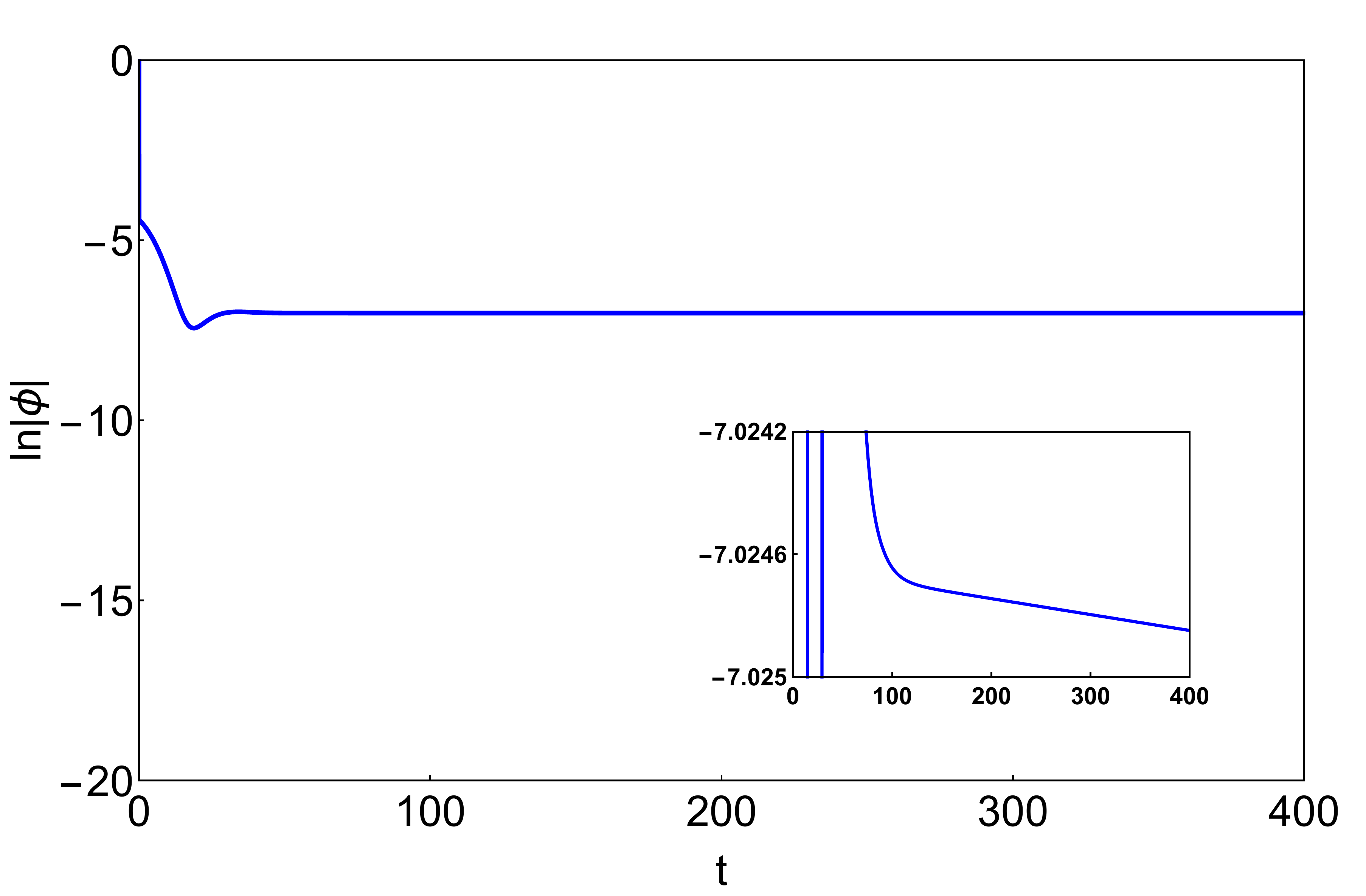}
\caption{Time evolution of the scalar perturbation $\phi$ non-minimally coupled to the Einstein tensor with coupling $\eta$ propagating on a fixed RNdS background. The BH mass is $M=1$ for all cases. {\bf Case 1 (upper left):} For $\Lambda=0.01$, $Q/Q_\text{max}=0.995$ and $\eta=-0.5$ the extracted dominant QNM belongs to the dS family with $l=1$, where $\omega=-0.0577i$. {\bf Case 2 (upper right):} For $\Lambda=0.01$, $Q/Q_\text{max}=0.999$ and $\eta=0.5$ the extracted dominant QNM belongs to the NE family with $l=0$, where $\omega=-0.0454i$. {\bf Case 3 (lower left):} For $\Lambda=0.06$, $Q/Q_\text{max}=0.995$ and $\eta=-0.5$ the extracted dominant QNM belongs to the PS family with $l=10$, where $\omega=2.1197-0.0767i$. {\bf Case 4 (lower right):} For $\Lambda=0.06$, $Q/Q_\text{max}=0.999$ and $\eta=0.5$ the extracted dominant QNM belongs to the NE family with $l=0$, where $\omega=-0.0422i$.}
\label{time evolutions}
\end{figure}

In Fig. \ref{time evolutions}, we show the evolution of a non-minimally coupled scalar perturbation on a RNdS background for different parameters. For Case 1, our frequency-domain calculations indicate that the dominant QNM will belong to the dS family, therefore it will be purely imaginary. By numerically integrating \eqref{eqmstar} with appropriate initial data and the designated parameters of Case 1, we can evolve the perturbation with respect to time. We see that, indeed, an exponential tail appears after the quasinormal ringing phase which dominates the late-time behavior of the ringdown signal\footnote{In Log-Linear scale, exponential functions are depicted by straight lines.}. By utilizing the Prony method, we have extracted the dominant mode at late times which matches very well the QNM calculated with the frequency-domain analysis.

For Cases 2 and 4, our frequency-domain calculations indicate that the dominant QNMs will belong to the NE family, therefore they are going to be purely imaginary, as well. By evolving the perturbation with respect to time, using the designated parameters of Cases 2 and 4, respectively, we see that, an exponential tail appears after the quasinormal ringing phase which dominates the late-time behavior of the ringdown signal. The extracted dominant QNMs at late times, again, match very well the QNMs calculated with the frequency-domain analysis. Although, in the corresponding panels of Fig. \ref{time evolutions}, the decay of the perturbation is not evident, due to the large decay timescale, by zooming we can still see that the perturbations indeed decay with respect to time.

Finally, for Case 3, the frequency-domain analysis indicates that the dominant QNM will belong to the PS family, therefore it is going to be complex. By evolving the perturbation with respect to time, using the designated parameters of Case 3, we see that the quasinormal ringing phase dominates the ringdown signal even at late times, due to the complex nature of the dominant QNM. The extracted dominant QNM at late times, again, match very well the QNM calculated with the frequency-domain analysis.

We complement our analysis with Table \ref{table}, where we demonstrate all dominant and a few subdominant QNMs for all cases discussed above. The table also reveals that the results of the time-domain and frequency-domain analysis, used here, match with very good precision.

\begin{table*}[ht]
\centering
\scalebox{0.8}{
\begin{tabular}{||c| c | c ||}
\hline
  \multicolumn{3}{||c||}{\bf{Case 1}} \\
   \hline
    $l$ & time$-$domain & frequency$-$domain \\ [0.5ex]
   \hline
    0 & -0.0928i  & -0.0925i  \\
   \hline
    1 & -0.0577i  &  -0.0577i \\
   \hline
   10 & 2.4787 - 0.0925i & 2.4771 - 0.0927i \\
   \hline\hline
   \multicolumn{3}{||c||}{\bf{Case 2}} \\
    \hline
     $l$ & time$-$domain & frequency$-$domain \\ [0.5ex]
   \hline
    0 & -0.0454i  & -0.0487i \\
    \hline
    1 &  -0.0577i & -0.0577i  \\
    \hline
    10 & 2.6524 - 0.0762i &  2.6505 - 0.0763i \\
    \hline
\end{tabular}}
\scalebox{0.8}{
\begin{tabular}{||c| c | c ||}
\hline
       \multicolumn{3}{||c||}{\bf{Case 3}} \\
        \hline
         $l$ & time$-$domain & frequency$-$domain \\ [0.5ex]
       \hline
        0 &  -0.0826i & -0.0826i \\
        \hline
         1 &  -0.1414i & -0.1414i \\
        \hline
        10 & 2.1197 - 0.0767i  & 2.1187 - 0.0768i \\
        \hline\hline
           \multicolumn{3}{||c||}{\bf{Case 4}} \\
            \hline
             $l$ & time$-$domain & frequency$-$domain \\ [0.5ex]
           \hline
            0 & -0.0422i &  -0.0422i\\
            \hline
                     1 &  -0.1414i & -0.1414i \\
            \hline
            10 & 2.2874 - 0.0644i  & 2.2862 - 0.0645i  \\
            \hline
\end{tabular}
}
\caption{QNMs of non-minimally coupled scalar perturbations propagating on a RNdS background. The modes have been extracted with a time-domain and a frequency-domain scheme. The cases depicted are: Case $1$, with $M=1$, $\Lambda=0.01$, $Q/Q_\text{max}=0.995$ and $\eta=-0.5$; Case $2$, with $M=1$, $\Lambda=0.01$, $Q/Q_\text{max}=0.999$ and $\eta=0.5$; Case $3$, with $M=1$, $\Lambda=0.06$, $Q/Q_\text{max}=0.995$ and $\eta=-0.5$ and, finally; Case 4, with $M=1$, $\Lambda=0.06$, $Q/Q_\text{max}=0.999$ and $\eta=0.5$.}
\label{table}
\end{table*}
\bibliography{SCC}

\providecommand{\href}[2]{#2}\begingroup\raggedright\begin{thebibliography}{10}

\bibitem{Penrose69}
R.~{Penrose}, \emph{{Gravitational Collapse: the Role of General Relativity}},
  {\emph{Nuovo Cimento Rivista Serie} {\bfseries 1} (1969) }.

\bibitem{Christodoulou:2008nj}
D.~Christodoulou, \emph{{The Formation of Black Holes in General Relativity}},
  pp.~24--34, 2008, \href{https://arxiv.org/abs/0805.3880}{{\ttfamily
  0805.3880}}, \href{https://doi.org/10.1142/9789814374552_0002}{DOI}.

\bibitem{Cardoso:2017soq}
V.~Cardoso, J.~L. Costa, K.~Destounis, P.~Hintz and A.~Jansen,
  \emph{{Quasinormal modes and Strong Cosmic Censorship}},
  \href{https://doi.org/10.1103/PhysRevLett.120.031103}{\emph{Phys. Rev. Lett.}
  {\bfseries 120} (2018) 031103}
  [\href{https://arxiv.org/abs/1711.10502}{{\ttfamily 1711.10502}}].

\bibitem{Luna:2018jfk}
R.~Luna, M.~Zilhão, V.~Cardoso, J.~L. Costa and J.~Natário, \emph{{Strong
  Cosmic Censorship: the nonlinear story}},
  \href{https://doi.org/10.1103/PhysRevD.99.064014}{\emph{Phys. Rev.}
  {\bfseries D99} (2019) 064014}
  [\href{https://arxiv.org/abs/1810.00886}{{\ttfamily 1810.00886}}].

\bibitem{Dafermos:2003wr}
M.~Dafermos, \emph{{The Interior of charged black holes and the problem of
  uniqueness in general relativity}}, {\emph{Commun. Pure Appl. Math.}
  {\bfseries 58} (2005) 0445}
  [\href{https://arxiv.org/abs/gr-qc/0307013}{{\ttfamily gr-qc/0307013}}].

\bibitem{Dafermos:2012np}
M.~Dafermos, \emph{{Black holes without spacelike singularities}},
  \href{https://doi.org/10.1007/s00220-014-2063-4}{\emph{Commun. Math. Phys.}
  {\bfseries 332} (2014) 729}
  [\href{https://arxiv.org/abs/1201.1797}{{\ttfamily 1201.1797}}].

\bibitem{BradyPoisson}
P.~R. {Brady} and E.~{Poisson}, \emph{{Cauchy horizon instability for
  Reissner-Nordstrom black holes in de Sitter space}},
  \href{https://doi.org/10.1088/0264-9381/9/1/011}{\emph{Classical and Quantum
  Gravity} {\bfseries 9} (1992) 121}.

\bibitem{Ori:1991zz}
A.~Ori, \emph{{Inner structure of a charged black hole: An exact mass-inflation
  solution}}, \href{https://doi.org/10.1103/PhysRevLett.67.789}{\emph{Phys.
  Rev. Lett.} {\bfseries 67} (1991) 789}.

\bibitem{Hintz:2016gwb}
P.~Hintz and A.~Vasy, \emph{{The global non-linear stability of the Kerr-de
  Sitter family of black holes}},
  \href{https://arxiv.org/abs/1606.04014}{{\ttfamily 1606.04014}}.

\bibitem{Hintz:2016jak}
P.~Hintz, \emph{Non-linear stability of the kerr - newman - de sitter family of
  charged black holes},
  \href{https://doi.org/10.1007/s40818-018-0047-y}{\emph{Annals of PDE} (2018)
  } [\href{https://arxiv.org/abs/1612.04489}{{\ttfamily 1612.04489}}].

\bibitem{Hartle}
S.~Chandrasekhar and J.~B. Hartle, \emph{On crossing the cauchy horizon of a
  reissner-nordstrom black-hole}, {\emph{Proceedings of the Royal Society of
  London. Series A, Mathematical and Physical Sciences} {\bfseries 384} (1982)
  301}.

\bibitem{Kokkotas:1999bd}
K.~D. Kokkotas and B.~G. Schmidt, \emph{{Quasinormal modes of stars and black
  holes}}, \href{https://doi.org/10.12942/lrr-1999-2}{\emph{Living Rev. Rel.}
  {\bfseries 2} (1999) 2}
  [\href{https://arxiv.org/abs/gr-qc/9909058}{{\ttfamily gr-qc/9909058}}].

\bibitem{Berti:2009kk}
E.~Berti, V.~Cardoso and A.~O. Starinets, \emph{{Quasinormal modes of black
  holes and black branes}},
  \href{https://doi.org/10.1088/0264-9381/26/16/163001}{\emph{Class. Quant.
  Grav.} {\bfseries 26} (2009) 163001}
  [\href{https://arxiv.org/abs/0905.2975}{{\ttfamily 0905.2975}}].

\bibitem{Konoplya:2011qq}
R.~A. Konoplya and A.~Zhidenko, \emph{{Quasinormal modes of black holes: From
  astrophysics to string theory}},
  \href{https://doi.org/10.1103/RevModPhys.83.793}{\emph{Rev. Mod. Phys.}
  {\bfseries 83} (2011) 793} [\href{https://arxiv.org/abs/1102.4014}{{\ttfamily
  1102.4014}}].

\bibitem{Hintz:2015jkj}
P.~Hintz and A.~Vasy, \emph{{Analysis of linear waves near the Cauchy horizon
  of cosmological black holes}},
  \href{https://doi.org/10.1063/1.4996575}{\emph{J. Math. Phys.} {\bfseries 58}
  (2017) 081509} [\href{https://arxiv.org/abs/1512.08004}{{\ttfamily
  1512.08004}}].

\bibitem{Dias:2018etb}
O.~J.~C. Dias, H.~S. Reall and J.~E. Santos, \emph{{Strong cosmic censorship:
  taking the rough with the smooth}},
  \href{https://doi.org/10.1007/JHEP10(2018)001}{\emph{JHEP} {\bfseries 10}
  (2018) 001} [\href{https://arxiv.org/abs/1808.02895}{{\ttfamily
  1808.02895}}].

\bibitem{Hod:2018dpx}
S.~Hod, \emph{{Strong cosmic censorship in charged black-hole spacetimes: As
  strong as ever}},
  \href{https://doi.org/10.1016/j.nuclphysb.2019.03.003}{\emph{Nucl. Phys.}
  {\bfseries B941} (2019) 636}
  [\href{https://arxiv.org/abs/1801.07261}{{\ttfamily 1801.07261}}].

\bibitem{Hod:2018lmi}
S.~Hod, \emph{{Quasinormal modes and strong cosmic censorship in near-extremal
  Kerr-Newman-de Sitter black-hole spacetimes}},
  \href{https://doi.org/10.1016/j.physletb.2018.03.020}{\emph{Phys. Lett.}
  {\bfseries B780} (2018) 221}
  [\href{https://arxiv.org/abs/1803.05443}{{\ttfamily 1803.05443}}].

\bibitem{Cardoso2}
V.~Cardoso, J.~L. Costa, K.~Destounis, P.~Hintz and A.~Jansen, \emph{{Strong
  cosmic censorship in charged black-hole spacetimes: still subtle}},
  \href{https://doi.org/10.1103/PhysRevD.98.104007}{\emph{Phys. Rev.}
  {\bfseries D98} (2018) 104007}
  [\href{https://arxiv.org/abs/1808.03631}{{\ttfamily 1808.03631}}].

\bibitem{Zhang1}
Y.~Mo, Y.~Tian, B.~Wang, H.~Zhang and Z.~Zhong, \emph{{Strong cosmic censorship
  for the massless charged scalar field in the Reissner-Nordstrom–de Sitter
  spacetime}}, \href{https://doi.org/10.1103/PhysRevD.98.124025}{\emph{Phys.
  Rev.} {\bfseries D98} (2018) 124025}
  [\href{https://arxiv.org/abs/1808.03635}{{\ttfamily 1808.03635}}].

\bibitem{Dias:2018ufh}
O.~J.~C. Dias, H.~S. Reall and J.~E. Santos, \emph{{Strong cosmic censorship
  for charged de Sitter black holes with a charged scalar field}},
  \href{https://doi.org/10.1088/1361-6382/aafcf2}{\emph{Class. Quant. Grav.}
  {\bfseries 36} (2019) 045005}
  [\href{https://arxiv.org/abs/1808.04832}{{\ttfamily 1808.04832}}].

\bibitem{Zhu:2014sya}
Z.~Zhu, S.-J. Zhang, C.~E. Pellicer, B.~Wang and E.~Abdalla, \emph{{Stability
  of Reissner-Nordström black hole in de Sitter background under charged
  scalar perturbation}}, \href{https://doi.org/10.1103/PhysRevD.90.044042,
  10.1103/PhysRevD.90.049904}{\emph{Phys. Rev.} {\bfseries D90} (2014) 044042}
  [\href{https://arxiv.org/abs/1405.4931}{{\ttfamily 1405.4931}}].

\bibitem{Konoplya:2014lha}
R.~A. Konoplya and A.~Zhidenko, \emph{{Charged scalar field instability between
  the event and cosmological horizons}},
  \href{https://doi.org/10.1103/PhysRevD.90.064048}{\emph{Phys. Rev.}
  {\bfseries D90} (2014) 064048}
  [\href{https://arxiv.org/abs/1406.0019}{{\ttfamily 1406.0019}}].

\bibitem{Destounis:2019hca}
K.~Destounis, \emph{{Superradiant instability of charged scalar fields in
  higher-dimensional Reissner-Nordstr\"om-de Sitter black holes}},
  \href{https://doi.org/10.1103/PhysRevD.100.044054}{\emph{Phys. Rev. D}
  {\bfseries 100} (2019) 044054}
  [\href{https://arxiv.org/abs/1908.06117}{{\ttfamily 1908.06117}}].

\bibitem{Ge:2018vjq}
B.~Ge, J.~Jiang, B.~Wang, H.~Zhang and Z.~Zhong, \emph{{Strong cosmic
  censorship for the massless Dirac field in the Reissner-Nordstrom-de Sitter
  spacetime}}, \href{https://doi.org/10.1007/JHEP01(2019)123}{\emph{JHEP}
  {\bfseries 01} (2019) 123}
  [\href{https://arxiv.org/abs/1810.12128}{{\ttfamily 1810.12128}}].

\bibitem{Destounis:2018qnb}
K.~Destounis, \emph{{Charged Fermions and Strong Cosmic Censorship}},
  \href{https://doi.org/10.1016/j.physletb.2019.06.015}{\emph{Phys. Lett.}
  {\bfseries B795} (2019) 211}
  [\href{https://arxiv.org/abs/1811.10629}{{\ttfamily 1811.10629}}].

\bibitem{Dias:2018ynt}
O.~J.~C. Dias, F.~C. Eperon, H.~S. Reall and J.~E. Santos, \emph{{Strong cosmic
  censorship in de Sitter space}},
  \href{https://doi.org/10.1103/PhysRevD.97.104060}{\emph{Phys. Rev.}
  {\bfseries D97} (2018) 104060}
  [\href{https://arxiv.org/abs/1801.09694}{{\ttfamily 1801.09694}}].

\bibitem{Rahman:2019uwf}
M.~Rahman, \emph{{On the validity of Strong Cosmic Censorship Conjecture in
  presence of Dirac fields}},
  \href{https://arxiv.org/abs/1905.06675}{{\ttfamily 1905.06675}}.

\bibitem{Liu:2019lon}
H.~Liu, Z.~Tang, K.~Destounis, B.~Wang, E.~Papantonopoulos and H.~Zhang,
  \emph{{Strong Cosmic Censorship in higher-dimensional Reissner-Nordström-de
  Sitter spacetime}},
  \href{https://doi.org/10.1007/JHEP03(2019)187}{\emph{JHEP} {\bfseries 03}
  (2019) 187} [\href{https://arxiv.org/abs/1902.01865}{{\ttfamily
  1902.01865}}].

\bibitem{Rahman:2018oso}
M.~Rahman, S.~Chakraborty, S.~SenGupta and A.~A. Sen, \emph{{Fate of Strong
  Cosmic Censorship Conjecture in Presence of Higher Spacetime Dimensions}},
  \href{https://doi.org/10.1007/JHEP03(2019)178}{\emph{JHEP} {\bfseries 03}
  (2019) 178} [\href{https://arxiv.org/abs/1811.08538}{{\ttfamily
  1811.08538}}].

\bibitem{Dafermos:2018tha}
M.~Dafermos and Y.~Shlapentokh-Rothman, \emph{{Rough initial data and the
  strength of the blue-shift instability on cosmological black holes with
  $\Lambda > 0$}}, \href{https://doi.org/10.1088/1361-6382/aadbcf}{\emph{Class.
  Quant. Grav.} {\bfseries 35} (2018) 195010}
  [\href{https://arxiv.org/abs/1805.08764}{{\ttfamily 1805.08764}}].

\bibitem{Gwak1}
B.~Gwak, \emph{{Strong Cosmic Censorship under Quasinormal Modes of
  Non-Minimally Coupled Massive Scalar Field}},
  \href{https://arxiv.org/abs/1812.04923}{{\ttfamily 1812.04923}}.

\bibitem{Guo:2019tjy}
H.~Guo, H.~Liu, X.-M. Kuang and B.~Wang, \emph{{Strong Cosmic Censorship in
  Charged de Sitter spacetime with Scalar Field Non-minimally Coupled to
  Curvature}},  \href{https://arxiv.org/abs/1905.09461}{{\ttfamily
  1905.09461}}.

\bibitem{Gan:2019jac}
Q.~Gan, G.~Guo, P.~Wang and H.~Wu, \emph{{Strong Cosmic Censorship for a Scalar
  Field in a Born-Infeld-de Sitter Black Hole}},
  \href{https://arxiv.org/abs/1907.04466}{{\ttfamily 1907.04466}}.

\bibitem{Horndeski1974}
G.~W. Horndeski, \emph{Second-order scalar-tensor field equations in a
  four-dimensional space},
  \href{https://doi.org/10.1007/BF01807638}{\emph{International Journal of
  Theoretical Physics} {\bfseries 10} (1974) 363}.

\bibitem{Nicolis:2008in}
A.~Nicolis, R.~Rattazzi and E.~Trincherini, \emph{{The Galileon as a local
  modification of gravity}},
  \href{https://doi.org/10.1103/PhysRevD.79.064036}{\emph{Phys. Rev.}
  {\bfseries D79} (2009) 064036}
  [\href{https://arxiv.org/abs/0811.2197}{{\ttfamily 0811.2197}}].

\bibitem{Deffayet:2009wt}
C.~Deffayet, G.~Esposito-Farese and A.~Vikman, \emph{{Covariant Galileon}},
  \href{https://doi.org/10.1103/PhysRevD.79.084003}{\emph{Phys. Rev.}
  {\bfseries D79} (2009) 084003}
  [\href{https://arxiv.org/abs/0901.1314}{{\ttfamily 0901.1314}}].

\bibitem{Deffayet:2009mn}
C.~Deffayet, S.~Deser and G.~Esposito-Farese, \emph{{Generalized Galileons: All
  scalar models whose curved background extensions maintain second-order field
  equations and stress-tensors}},
  \href{https://doi.org/10.1103/PhysRevD.80.064015}{\emph{Phys. Rev.}
  {\bfseries D80} (2009) 064015}
  [\href{https://arxiv.org/abs/0906.1967}{{\ttfamily 0906.1967}}].

\bibitem{Papantonopoulos:2019eff}
E.~Papantonopoulos, \emph{{Effects of the kinetic coupling of matter to
  curvature}}, \href{https://doi.org/10.1142/S0218271819420070}{\emph{Int. J.
  Mod. Phys.} {\bfseries D28} (2019) 1942007}.

\bibitem{Sushkov:2009hk}
S.~V. Sushkov, \emph{{Exact cosmological solutions with nonminimal derivative
  coupling}}, \href{https://doi.org/10.1103/PhysRevD.80.103505}{\emph{Phys.
  Rev.} {\bfseries D80} (2009) 103505}
  [\href{https://arxiv.org/abs/0910.0980}{{\ttfamily 0910.0980}}].

\bibitem{Germani:2010gm}
C.~Germani and A.~Kehagias, \emph{{New Model of Inflation with Non-minimal
  Derivative Coupling of Standard Model Higgs Boson to Gravity}},
  \href{https://doi.org/10.1103/PhysRevLett.105.011302}{\emph{Phys. Rev. Lett.}
  {\bfseries 105} (2010) 011302}
  [\href{https://arxiv.org/abs/1003.2635}{{\ttfamily 1003.2635}}].

\bibitem{Koutsoumbas:2015ekk}
G.~Koutsoumbas, K.~Ntrekis, E.~Papantonopoulos and M.~Tsoukalas,
  \emph{{Gravitational Collapse of a Homogeneous Scalar Field Coupled
  Kinematically to Einstein Tensor}},
  \href{https://doi.org/10.1103/PhysRevD.95.044009}{\emph{Phys. Rev.}
  {\bfseries D95} (2017) 044009}
  [\href{https://arxiv.org/abs/1512.05934}{{\ttfamily 1512.05934}}].

\bibitem{Kolyvaris:2017efz}
T.~Kolyvaris and E.~Papantonopoulos, \emph{{Superradiant Amplification of a
  Scalar Wave Coupled Kinematically to Curvature Scattered off a
  Reissner-Nordstr\"om Black Hole}},
  \href{https://arxiv.org/abs/1702.04618}{{\ttfamily 1702.04618}}.

\bibitem{Kolyvaris:2018zxl}
T.~Kolyvaris, M.~Koukouvaou, A.~Machattou and E.~Papantonopoulos,
  \emph{{Superradiant instabilities in scalar-tensor Horndeski theory}},
  \href{https://doi.org/10.1103/PhysRevD.98.024045}{\emph{Phys. Rev.}
  {\bfseries D98} (2018) 024045}
  [\href{https://arxiv.org/abs/1806.11110}{{\ttfamily 1806.11110}}].

\bibitem{Minamitsuji:2014hha}
M.~Minamitsuji, \emph{{Black hole quasinormal modes in a scalar-tensor theory
  with field derivative coupling to the Einstein tensor}},
  \href{https://doi.org/10.1007/s10714-014-1785-0}{\emph{Gen. Rel. Grav.}
  {\bfseries 46} (2014) 1785}
  [\href{https://arxiv.org/abs/1407.4901}{{\ttfamily 1407.4901}}].

\bibitem{Yu:2018zqd}
S.~Yu and C.~Gao, \emph{{Quansinormal modes of static and spherically symmetric
  black holes with the derivative coupling}},
  \href{https://doi.org/10.1007/s10714-019-2500-y}{\emph{Gen. Rel. Grav.}
  {\bfseries 51} (2019) 16} [\href{https://arxiv.org/abs/1807.05024}{{\ttfamily
  1807.05024}}].

\bibitem{Konoplya:2018qov}
R.~A. Konoplya, Z.~Stuchlík and A.~Zhidenko, \emph{{Massive nonminimally
  coupled scalar field in Reissner-Nordström spacetime: Long-lived quasinormal
  modes and instability}},
  \href{https://doi.org/10.1103/PhysRevD.98.104033}{\emph{Phys. Rev.}
  {\bfseries D98} (2018) 104033}
  [\href{https://arxiv.org/abs/1808.03346}{{\ttfamily 1808.03346}}].

\bibitem{Abdalla:2018ggo}
E.~Abdalla, B.~Cuadros-Melgar, J.~de~Oliveira, A.~B. Pavan and C.~E. Pellicer,
  \emph{{Vectorial and spinorial perturbations in Galileon Black Holes:
  Quasinormal modes, quasiresonant modes and stability}},
  \href{https://doi.org/10.1103/PhysRevD.99.044023}{\emph{Phys. Rev.}
  {\bfseries D99} (2019) 044023}
  [\href{https://arxiv.org/abs/1810.01198}{{\ttfamily 1810.01198}}].

\bibitem{Fontana:2018fof}
R.~D.~B. Fontana, J.~de~Oliveira and A.~B. Pavan, \emph{{Dynamical evolution of
  non-minimally coupled scalar field in spherically symmetric de Sitter
  spacetimes}},
  \href{https://doi.org/10.1140/epjc/s10052-019-6831-3}{\emph{Eur. Phys. J.}
  {\bfseries C79} (2019) 338}
  [\href{https://arxiv.org/abs/1808.01044}{{\ttfamily 1808.01044}}].

\bibitem{Abdalla:2019irr}
E.~Abdalla, B.~Cuadros-Melgar, R.~D.~B. Fontana, J.~de~Oliveira,
  E.~Papantonopoulos and A.~B. Pavan, \emph{{Instability of a
  Reissner-Nordström-AdS black hole under perturbations of a scalar field
  coupled to the Einstein tensor}},
  \href{https://doi.org/10.1103/PhysRevD.99.104065}{\emph{Phys. Rev.}
  {\bfseries D99} (2019) 104065}
  [\href{https://arxiv.org/abs/1903.10850}{{\ttfamily 1903.10850}}].

\bibitem{Costa:2017tjc}
J.~L. Costa, P.~M. Girão, J.~Natário and J.~D. Silva, \emph{{On the
  Occurrence of Mass Inflation for the Einstein–Maxwell-Scalar Field System
  with a Cosmological Constant and an Exponential Price Law}},
  \href{https://doi.org/10.1007/s00220-018-3122-z}{\emph{Commun. Math. Phys.}
  {\bfseries 361} (2018) 289}
  [\href{https://arxiv.org/abs/1707.08975}{{\ttfamily 1707.08975}}].

\bibitem{Dafermos:2017dbw}
M.~Dafermos and J.~Luk, \emph{{The interior of dynamical vacuum black holes I:
  The $C^0$-stability of the Kerr Cauchy horizon}},
  \href{https://arxiv.org/abs/1710.01722}{{\ttfamily 1710.01722}}.

\bibitem{Wald:1984rg}
R.~M. Wald, \emph{{General Relativity}}. Chicago Univ. Pr., Chicago, USA, 1984,
  \href{https://doi.org/10.7208/chicago/9780226870373.001.0001}{10.7208/chicago/9780226870373.001.0001}.

\bibitem{Faraoni:2015sja}
V.~Faraoni, \emph{{Quasilocal energy in modified gravity}},
  \href{https://doi.org/10.1088/0264-9381/33/1/015007}{\emph{Class. Quant.
  Grav.} {\bfseries 33} (2016) 015007}
  [\href{https://arxiv.org/abs/1508.06849}{{\ttfamily 1508.06849}}].

\bibitem{Szabados2009}
L.~B. Szabados, \emph{Quasi-local energy-momentum and angular momentum in
  general relativity}, \href{https://doi.org/10.12942/lrr-2009-4}{\emph{Living
  Reviews in Relativity} {\bfseries 12} (2009) 4}.

\bibitem{Papallo:2017qvl}
G.~Papallo and H.~S. Reall, \emph{{On the local well-posedness of Lovelock and
  Horndeski theories}},
  \href{https://doi.org/10.1103/PhysRevD.96.044019}{\emph{Phys. Rev.}
  {\bfseries D96} (2017) 044019}
  [\href{https://arxiv.org/abs/1705.04370}{{\ttfamily 1705.04370}}].

\bibitem{Papallo:2017ddx}
G.~Papallo, \emph{{On the hyperbolicity of the most general Horndeski theory}},
  \href{https://doi.org/10.1103/PhysRevD.96.124036}{\emph{Phys. Rev.}
  {\bfseries D96} (2017) 124036}
  [\href{https://arxiv.org/abs/1710.10155}{{\ttfamily 1710.10155}}].

\bibitem{LeFloch:2014zva}
P.~G. LeFloch and Y.~Ma, \emph{{Mathematical Validity of the f(R) Theory of
  Modified Gravity}},  \href{https://arxiv.org/abs/1412.8151}{{\ttfamily
  1412.8151}}.

\bibitem{Jansen:2017oag}
A.~Jansen, \emph{{Overdamped modes in Schwarzschild-de Sitter and a Mathematica
  package for the numerical computation of quasinormal modes}},
  \href{https://doi.org/10.1140/epjp/i2017-11825-9}{\emph{Eur. Phys. J. Plus}
  {\bfseries 132} (2017) 546}
  [\href{https://arxiv.org/abs/1709.09178}{{\ttfamily 1709.09178}}].

\bibitem{Dias:2015nua}
O.~J.~C. Dias, J.~E. Santos and B.~Way, \emph{{Numerical Methods for Finding
  Stationary Gravitational Solutions}},
  \href{https://doi.org/10.1088/0264-9381/33/13/133001}{\emph{Class. Quant.
  Grav.} {\bfseries 33} (2016) 133001}
  [\href{https://arxiv.org/abs/1510.02804}{{\ttfamily 1510.02804}}].

\bibitem{Gundlach:1993tp}
C.~Gundlach, R.~H. Price and J.~Pullin, \emph{{Late time behavior of stellar
  collapse and explosions: 1. Linearized perturbations}},
  \href{https://doi.org/10.1103/PhysRevD.49.883}{\emph{Phys. Rev.} {\bfseries
  D49} (1994) 883} [\href{https://arxiv.org/abs/gr-qc/9307009}{{\ttfamily
  gr-qc/9307009}}].

\bibitem{Berti:2007dg}
E.~Berti, V.~Cardoso, J.~A. Gonzalez and U.~Sperhake, \emph{{Mining information
  from binary black hole mergers: A Comparison of estimation methods for
  complex exponentials in noise}},
  \href{https://doi.org/10.1103/PhysRevD.75.124017}{\emph{Phys. Rev.}
  {\bfseries D75} (2007) 124017}
  [\href{https://arxiv.org/abs/gr-qc/0701086}{{\ttfamily gr-qc/0701086}}].

\bibitem{Schutz:1985zz}
B.~F. Schutz and C.~M. Will, \emph{{Black hole normal modes: A semianalytic
  approach}}, \href{https://doi.org/10.1086/184453}{\emph{Astrophys. J.}
  {\bfseries 291} (1985) L33}.

\bibitem{Brill:1993tw}
D.~R. Brill and S.~A. Hayward, \emph{{Global structure of a black hole cosmos
  and its extremes}},
  \href{https://doi.org/10.1088/0264-9381/11/2/008}{\emph{Class. Quant. Grav.}
  {\bfseries 11} (1994) 359}
  [\href{https://arxiv.org/abs/gr-qc/9304007}{{\ttfamily gr-qc/9304007}}].

\bibitem{Rendall:2003ks}
A.~D. Rendall, \emph{{Asymptotics of solutions of the Einstein equations with
  positive cosmological constant}},
  \href{https://doi.org/10.1007/s00023-004-0189-1}{\emph{Annales Henri
  Poincare} {\bfseries 5} (2004) 1041}
  [\href{https://arxiv.org/abs/gr-qc/0312020}{{\ttfamily gr-qc/0312020}}].

\bibitem{Du:2004jt}
D.~P. Du, B.~Wang and R.~K. Su, \emph{{Quasinormal modes in pure de Sitter
  space-times}}, \href{https://doi.org/10.1103/PhysRevD.70.064024}{\emph{Phys.
  Rev.} {\bfseries D70} (2004) 064024}
  [\href{https://arxiv.org/abs/hep-th/0404047}{{\ttfamily hep-th/0404047}}].

\bibitem{LopezOrtega:2006my}
A.~Lopez-Ortega, \emph{{Quasinormal modes of D-dimensional de Sitter
  spacetime}}, \href{https://doi.org/10.1007/s10714-006-0335-9}{\emph{Gen. Rel.
  Grav.} {\bfseries 38} (2006) 1565}
  [\href{https://arxiv.org/abs/gr-qc/0605027}{{\ttfamily gr-qc/0605027}}].

\bibitem{VasydS}
A.~Vasy, \emph{The wave equation on asymptotically de {S}itter-like spaces},
  {\emph{Advances in Mathematics} {\bfseries 223} (2010) 49}.

\bibitem{Klainerman:2012wt}
S.~Klainerman, I.~Rodnianski and J.~Szeftel, \emph{{The Bounded L2 Curvature
  Conjecture}},  \href{https://arxiv.org/abs/1204.1767}{{\ttfamily 1204.1767}}.

\bibitem{2011ApJ...736..103B}
L.~W. {Brenneman}, C.~S. {Reynolds}, M.~A. {Nowak}, R.~C. {Reis}, M.~{Trippe},
  A.~C. {Fabian} et~al., \emph{{The Spin of the Supermassive Black Hole in NGC
  3783}}, \href{https://doi.org/10.1088/0004-637X/736/2/103}{\emph{ApJ}
  {\bfseries 736} (2011) 103}
  [\href{https://arxiv.org/abs/1104.1172}{{\ttfamily 1104.1172}}].

\bibitem{FastSpin}
G.~Risaliti, F.~Harrison, K.~Madsen, D.~Walton, S.~Boggs, F.~Christensen
  et~al., \emph{A rapidly spinning supermassive black hole at the centre of ngc
  1365}, {\emph{Nature} {\bfseries 494} (2013) 449}.

\end{thebibliography}\endgroup

\end{document}